\documentclass[11pt,a4paper]{article}
\usepackage{jheppub}
\usepackage[english]{babel}
\usepackage{amssymb,amsmath}
\usepackage{mathrsfs}
\usepackage{hyperref}
\usepackage[dvipsnames,x11names]{xcolor}
\usepackage{graphicx}
\usepackage{subfig}
\usepackage{mathtools}
\usepackage{slashed}
\usepackage{float}
\usepackage[normalem]{ulem}

\makeatletter
\gdef\@fpheader{}
\makeatother

\usepackage{dcolumn}
\usepackage{textcomp}
\usepackage[dvipsnames]{xcolor}
\usepackage{xfrac}
\usepackage{slashed}
\usepackage{multirow}

\def\be{\begin{equation}}
\def\ee{\end{equation}}
\def\figs/B{B}
\def\bea{\begin{eqnarray}}
\def\eea{\end{eqnarray}}
\def\bg{\begin{eqnarray}}
\def\nd{\end{eqnarray}}

\def\cos{{\rm cos}}

\def\log{{\rm log}}
\def\ln{{\rm log}}

\newcommand{\ca}{n}

\def\beq{\begin{equation}}
\def\eeq{\end{equation}}

\newcommand{\intd}{\int {\rm d}}

\begin{document}

\preprint{MIT-CTP/5860}

\title{Precision Unitarity Calculations in Inflationary Models} 

\author[a,b]{Thomas Steingasser,}
\author[c]{Mark P. Hertzberg,}
\author[a]{and David I.~Kaiser}

\affiliation[a]{%
Department of Physics, Massachusetts Institute of Technology, Cambridge, MA 02139, USA }

\affiliation[b]{Black Hole Initiative at Harvard University, 20 Garden Street, Cambridge, MA 02138, USA
}%

\affiliation[c]{%
Institute of Cosmology, Department of Physics and Astronomy, Tufts University, Medford, MA 02155, USA
}%

 \emailAdd{tstngssr@mit.edu}
 \emailAdd{mark.hertzberg@tufts.edu}
 \emailAdd{dikaiser@mit.edu}

\date{\today}

\abstract{
We revisit perturbative unitarity in scalar field inflation with a nonminimal coupling, with Higgs inflation serving as the most prominent example. Although such models are phenomenologically successful, it is critical to examine whether or not unitarity violations spoil their theoretical self-consistency. The analysis of these issues has so far typically relied on order-of-magnitude estimates of scattering amplitudes, which are appropriate for generic parameters. It is not evident that these methods apply to scenarios relying on a near-critical inflationary potential, for which an interplay of both small scalar self-couplings and nonminimal couplings could partially alleviate the unitarity issues. To allow for an exploration of this possibility, we consider the full $S$-matrix for the relevant scattering processes, taking into account important phase space volume factors, leading to a precise evaluation of the cut-off scale. In the single-field case, we demonstrate that near-criticality raises the cut-off scale considerably, compared to previous estimates. In the multifield case, momentum-dependent self-interactions in the kinetic sector lower the cut-off compared to the single-field case to a value comparable to but slightly larger than previous estimates. We carefully study both the single-field and multifield cases in metric and metric-affine (Palatini) formulations of gravity, 
as well as introduce a new phenomenologically viable model with a canonical kinetic term and a significantly raised cut-off, and discuss the importance of background field effects.}

\maketitle

\section{Introduction}\label{sec:intro}

Modern fundamental physics rests on two pillars. At large scales, the Universe is described with good accuracy by the $\Lambda$CDM model, while the physics of small scales is captured remarkably well by the Standard Model of particle physics (SM), which has been completed with the discovery of the Higgs boson at the LHC~\cite{ATLAS:2012yve,CMS:2012qbp}. One of the main challenges of both particle physics and cosmology remains to find microphysical explanations for the ``ingredients'' of the $\Lambda$CDM model, such as the mechanism responsible for early-universe inflation. An economical possibility, known as ``Higgs inflation,'' is to identify the SM Higgs field as the inflaton~\cite{Bezrukov:2007ep}---after all, the Higgs field is now known to exist, and its properties have been measured to high accuracy. (See Ref.~\cite{Rubio:2018ogq} for a review of Higgs inflation and its variants.) The original Higgs inflation model incorporated a nonminimal coupling of the Higgs field ${\bf H}$ to the spacetime Ricci curvature scalar $R$ of the form $\xi {\bf H}^\dagger {\bf H} R$; such couplings arise inevitably for any self-interacting scalar fields in a curved spacetime due to radiative effects~\cite{Birrell:1982ix,Buchbinder:1992rb,Parker:2009uva}. 

Early iterations of this scenario, however, faced a significant challenge: to match the measured amplitude of temperature anisotropies in the cosmic microwave background radiation (CMB), the Higgs quartic self-coupling $\lambda$ and its nonminimal coupling to gravity $\xi$ must satisfy $\lambda  / \xi^2 \simeq 10^{-9}$. For $\lambda \sim {\cal O} (10^{-1})$, this would require an enormous value of the dimensionless nonminimal coupling constant, $\xi \sim {\cal O} (10^4)$~\cite{Bezrukov:2007ep}. Such a large value not only raised the question of naturalness but has also been argued to cause a breakdown of perturbative unitarity at scales larger than~\cite{Burgess:2009ea,Burgess:2010zq,Barbon:2009ya,Hertzberg:2010dc,Giudice:2010ka,Lerner:2011it,Barbon:2015fla} 
\begin{equation}\label{eq:LambdaUVwiggle}
    E_{\rm max} \sim \frac{M_{\rm Pl}}{\xi},
\end{equation}
where $M_{\rm Pl} \equiv 1 / \sqrt{ 8 \pi G} = 2.43 \cdot 10^{18} \, {\rm GeV}$ is the reduced Planck mass. This breakdown arises from scattering amplitudes of particles around the global minimum of the potential. 

A cut-off at some scale $E_{\rm max}$ implies the existence of new physics (NP) at a scale $\Lambda_{\rm NP}<E_{\rm max}$ capable of restoring unitarity. Crucially, for $E_{\rm max}$ in Eq.~(\ref{eq:LambdaUVwiggle}), such NP would inevitably also influence the inflationary dynamics, making predictions for cosmological observables highly dependent on the details of its dynamics and couplings to the Higgs field. A concrete example for this behavior relying on an additional scalar field was given in Ref.~\cite{Barbon:2015fla}, where it was shown that it is indeed this new field that drives inflation rather than the Higgs. In Sec.~\ref{sec:unitarity}, we provide an extensive discussion of the connection between unitarity and the scale of NP more generally.

Such large values for $\xi$ can be avoided by taking into account the running of $\lambda$, which allows for an intriguing possibility known as ``critical Higgs inflation.'' Within the 95\% confidence ellipsoid of the relevant SM parameters there exists a region in which the quartic Higgs coupling develops a minimum with $\lambda\sim \mathcal{O}(10^{-5})$ at a large RG scale $\mu_*\sim \mathcal{O}(10^{-1}) \, M_{\rm Pl}$. This allows two-loop corrections to $\lambda$ to give rise to the formation of a near-inflection point in the effective potential for the Higgs field. (For recent studies of the relevant RG trajectories, see Refs.~\cite{Steingasser:2023ugv,Steingasser:2024hqi,Detering:2024vxs,Hiller:2024zjp}.) This feature has long been known to be capable of giving rise to successful slow-roll inflation, with a value of the nonminimal coupling $\xi \sim {\cal O} (30)$~\cite{Bezrukov:2014bra}---and thus, a significantly higher scale of unitarity violation. Although this effect is insufficient to push $\Lambda_{\rm NP}$ above the scales relevant during inflation, it is indeed sufficient to bring them close to one another. In this regime, numerical prefactors of $\mathcal{O}(1-10)$ neglected in the earlier investigations leading to the bound in Eq.~\eqref{eq:LambdaUVwiggle} become important. 

We carefully compute the cut-off scale $E_{\rm max}$ through a detailed analysis of scattering amplitudes, both in the single field and multifield cases. In Sec.~\ref{sec:EFT}, we review in detail how the nonminimal coupling can be incorporated in a simple Effective Field Theory (EFT) language, making manifest its impact on the scattering of scalar particles. In Sec.~\ref{sec:UnitarityBounds}, we first review the connection between unitarity and scattering amplitudes on the most general level before providing our results for the scale of unitarity violation, separately considering scatterings arising from interactions through the potential and momentum-dependent scatterings linked to a non-diagonal field-space metric in multifield models. For the latter case, our results agree parametrically with those in the literature. For the first case, we find a dependence of the unitarity violation scale on the relevant coupling constant, such that a weak coupling corresponds to a higher scale of unitarity violation. We then briefly revisit the problem in Palatini gravity, where we find a similar behavior. To illustrate the effect of the interactions arising from the field-space metric we furthermore introduce a toy model with canonically normalized kinetic terms, for which we find that unitarity is restored up to significantly higher energies---especially for the case of a small self-interaction. To put all these results into context, in Sec.~\ref{sec:InflationaryObservables} we provide simple estimates based on the evolution of the relevant fields during inflation in the slow-roll approximation.

It is well-known that the value of $E_{\rm max}$ depends on the background around which the model is expanded, with the inflationary background generally leading to a parametrically larger value for $E_{\rm max}\sim M_{\rm Pl}/\sqrt{\xi}$~\cite{Bezrukov:2010jz,Bezrukov:2012hx,Bezrukov:2013fka,Calmet:2013hia,MagninDissertation}. This phenomenon has been argued to be mostly inconsequential for theories expected to be reliable both during inflation and at exponentially smaller energies, as the smaller value $\Lambda_{\rm NP}$  inferred by expanding around the low-energy minimum would still be smaller than the field values during inflation~\cite{Burgess:2009ea,Burgess:2010zq,Barbon:2009ya,Hertzberg:2010dc,Giudice:2010ka,Lerner:2011it,Barbon:2015fla}. However, this would no longer be the case if the model were only required to be viable during inflation. While it lies beyond the scope of our analysis to decide whether or not a self-consistent theory can be constructed whose high-energy degrees of freedom can be described by a Lagrangian with the same structure as considered in this article, we can nevertheless investigate the compliance of this possibility with unitarity. Thus, in Sec.~\ref{sec:UnitarityInflation}, we review the derivation of $E_{\rm max}$ when expanding around the inflationary background for the special case of the Lagrangian associated with Higgs inflation. We find that $E_{\rm max}$ is indeed raised to be of order of the field values during inflation, albeit marginally smaller.

In Appendix~\ref{appendixOnShellQuanta}, we demonstrate that the number of on-shell quanta with $E \sim E_{\rm max}$ within a given Hubble sphere, which would be available to participate in the types of high-energy scattering that yield the unitarity cut-off scale $E_{\rm max}$, is exponentially suppressed both during and after inflation.

Throughout this article, we consider scattering processes involving scalar particles at tree level. For single-field models with non-polynomial effective potentials---such as Higgs inflation in unitary gauge, when treated in the Einstein frame---it has been argued that unitarity may be preserved via ``self-healing,'' \cite{Calmet:2013hia,MagninDissertation}, upon resumming classes of Feynman diagrams to all orders. (See also Ref.~\cite{Aydemir:2012nz}.) Since we find that the most stringent bounds on the unitarity cut-off scale $E_{\rm max}$ come from tree-level scattering in multifield models involving the momentum-dependent operators associated with the noncanonical kinetic sector---rather than the non-polynomial effective potential---we leave possible implications of resummations beyond tree-level for future work.

\section{Unitarity violations and new physics in effective field theories}\label{sec:unitarity}

EFTs have proven to be remarkably powerful tools with which to treat the dynamics of complicated systems and extract precise predictions for observables within bounded regimes of validity. EFTs typically rely on a hierarchy of scales, such that high-energy (or short-distance) processes can be integrated out of the effective action when aiming to provide an accurate description of lower-energy (or longer-distance) phenomena. (For detailed reviews, see, e.g., Refs.~\cite{Burgess:2007pt,Burgess:2020tbq}.)

In general, there need not exist a one-to-one mapping between the relevant degrees of freedom treated by an EFT within a specific energy range and the relevant degrees of freedom at arbitrarily different energy scales, well beyond the self-consistent regime of validity for a given EFT~\cite{Burgess:2007pt,Burgess:2020tbq}. When calculating the scale at which perturbative unitarity is violated for an EFT---and hence the scale at which NP would be required---one must select the relevant background around which to expand the effective action, since the Feynman rules for scattering processes whose amplitudes contribute to the $S$-matrix depend on the form of the effective action. When considering an EFT that should reduce to the Standard Model at the electroweak symmetry-breaking scale, for example, one typically expands the effective action around the vacuum expectation value of the relevant fields in Minkowski spacetime. (For reviews see, e.g., Refs.~\cite{Brivio:2017vri,Isidori:2023pyp,Logan:2022uus}.) On the other hand, when considering an EFT for inflation one expands the effective action around the nonzero inflaton condensate $\langle \phi \rangle =\phi_c$ and the nonzero expansion rate $H$ of the quasi-de Sitter background spacetime \cite{Cheung:2007st,Weinberg:2008hq,Senatore:2010wk,Pinol:2024arz,Azhar:2018nol,Azhar:2022yip}. 

A special case involves theories that are supposed to provide a reliable description across a wide range of scales, the most prominent example for which is Higgs inflation~\cite{Bezrukov:2007ep,Rubio:2018ogq}, in which the Higgs field serves as both the inflaton at near-Planckian scales while also being responsible for spontaneous symmetry breaking near the exponentially smaller electroweak scale. By identifying the SM Higgs field as the {\it same} dynamical degree of freedom across these vastly different energy scales, the question of unitarity becomes relevant {\it both} around field configurations $\phi \sim 0$ as well as around inflationary configurations $\phi \sim \phi_c$. (See also Ref.~\cite{Weymann-Despres:2023wly}.) Therefore, in this article, we calculate the unitarity violation scale based on scattering amplitudes around both $\phi \sim 0$ and $\phi \sim \phi_c$. 

As laid out in detail below, upon performing a conformal transformation to rescale the spacetime metric $g_{\mu\nu} (x)$, a nonminimal coupling between a scalar field and the spacetime Ricci scalar can be understood as inducing an infinite tower of higher-dimensional operators involving that scalar field. In multifield models, two distinct sets of higher-dimensional operators become important: those arising from the tree-level potential $V (\Phi^I)$, and momentum-dependent operators associated with the noncanonical kinetic terms\\ ${\cal G}_{IJ} (\Phi^K) \, \partial_\mu \Phi^I \,\partial_\nu \Phi^J$. In Sec.~\ref{sec:EFT} we identify the unitarity cut-off scale for scattering around a vanishing Higgs vacuum expectation value in Minkowski spacetime, $\phi \sim 0$. We calculate the scattering amplitudes under various assumptions about the underlying gravitational theory (metric versus metric-affine or Palatini), for both single-field and multifield cases. Later, in Sec.~\ref{sec:UnitarityInflation}, we repeat the exercise for scattering amplitudes expanded around the inflationary configuration, $\phi \sim \phi_c$. 

\begin{figure}[h!]
    \centering
    \includegraphics[width=0.99 \textwidth]{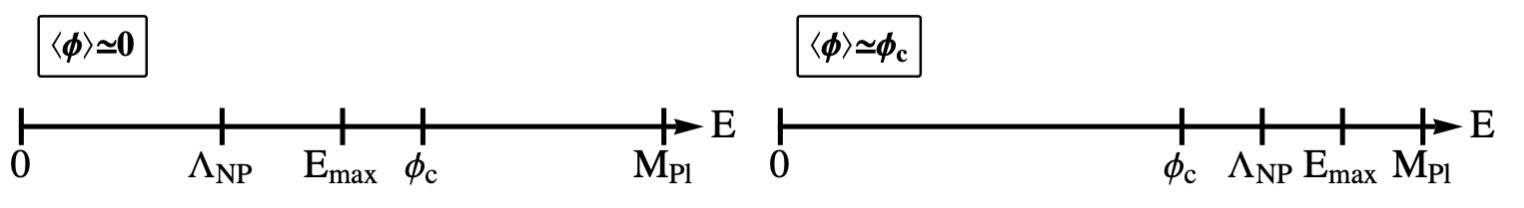}
    \caption{\textit{Left:} The ordering of scales implied by scattering calculations around a vanishing vacuum expectation value, $\langle \phi \rangle \simeq 0$. The breakdown of unitarity at the scale $E_{\rm max}$ implies the need for some NP to become relevant at some smaller energy scale $\Lambda_{\rm NP}$. In cases like Higgs inflation in which the same degrees of freedom appear in the effective descriptions during inflation and at exponentially lower energy scales, the scale  $\Lambda_{\rm NP}$ is smaller than the field values $\phi_c$ during inflation, making the classical dynamics of the inflation field in principle sensitive to this NP. \textit{Right:} During inflation the scalar field takes a large value, raising the value of $E_{\rm max}$. If this effect is strong enough, and if one drops the requirement for the effective theory to reliably describe the regime $\langle \phi \rangle\simeq 0$ as well as $\langle \phi \rangle \simeq \phi_c$, then such an EFT can accurately describe the inflationary dynamics in a self-consistent manner.}
    \label{fig:scales}
\end{figure}

In each case, the higher-order terms in the effective action allow for tree-level scattering processes of the form $2 \phi \to N \phi$ (suppressing species indices for now). The amplitudes of these processes become large enough to violate unitarity for center-of-mass energies $E> E_{\rm max}$. This implies the existence of NP at some scale $\Lambda_{\rm NP}< E_{\rm max}$ capable of restoring unitarity. This observation has proven especially important in the context of Higgs inflation: when considering scattering processes of Higgs particles around the global minimum $\langle \phi \rangle \simeq 0$, the resulting scale $\Lambda_{\rm NP}$ is generally \textit{smaller} than the values of the scalar fields during inflation, $\phi_c$. See Fig.~\ref{fig:scales}. This implies a significant sensitivity of any inflationary predictions on this NP, undermining the predictability of the model based on the stipulated relationship between relevant degrees of freedom at both $\phi_c \sim {\cal O} (10^{17} \> {\rm GeV} )$ and at the electroweak scale $v_{\rm EW} \sim {\cal O}(10^2 \> {\rm GeV})$. (See, e.g., Refs.~\cite{Burgess:2009ea,Burgess:2010zq,Barbon:2009ya,Giudice:2010ka,Lerner:2011it,Barbon:2015fla}.) This sensitivity can manifest through such effects as a change in the running of the couplings, additional terms in the potential, or the transition to a higher-dimensional field space~\cite{Barbon:2015fla}.

As has been recognized within the literature on Higgs inflation for some time (see, e.g., Refs.~\cite{Bezrukov:2010jz,Bezrukov:2012hx,Bezrukov:2013fka,Calmet:2013hia,MagninDissertation}), when computed in the background of a sufficiently large inflaton condensate $\langle \phi \rangle \simeq \phi_c$, $E_{\rm max}$ can be raised parametrically relative to the value obtained around $\langle \phi \rangle \simeq 0$. In particular, this effect can be strong enough to lift $E_{\rm max}$ above $\phi_c$. For models such as Higgs inflation, in which the same EFT is meant to accurately describe physics {\it both} around $\langle \phi \rangle \simeq 0$ {\it and} around $\langle \phi \rangle \simeq \phi_c$, the observation that $E_{\rm max} > \phi_c$ during inflation appears to be inconsequential: the hierarchy $\Lambda_{\rm NP} < E_{\rm max} < \phi_c$, arising from scattering amplitudes around $\langle \phi \rangle \simeq 0$, would imply the need for some (unknown) NP at scales relevant for inflation. On the other hand, if a Higgs-like inflationary model were treated as an EFT for inflationary dynamics, but \textit{not} assumed to describe physics at arbitrarily lower energy scales, then cases in which $E_{\rm max} > \phi_c$ at inflationary scales would provide a self-consistent description of the inflationary dynamics. For the case of vanilla Higgs inflation, we find in Sec.~\ref{sec:UnitarityInflation} that this condition is (marginally) violated even when computing scattering amplitudes around $\langle \phi \rangle \simeq \phi_c$.

\section{Gravity with nonminimal coupling as an effective field theory}\label{sec:EFT}

We consider a real, $W$-component scalar field $\Phi^I = (\phi^1, ... , \phi^W)$ with nonminimal coupling $\xi$ to gravity and $SO(W)$ symmetry among the scalars. Whenever we perform calculations that depend on the value of $W$, we choose the special case $W=4$, corresponding, e.g., to the Higgs doublet. The action of such a model is given by
\begin{equation}
    S = \intd^4 x \sqrt{- \bar{g}} \left[  f(\Phi) \bar{R} - \frac{1}{2} \delta_{IJ} \bar{g}^{\mu\nu} \partial_\mu \Phi^I \partial_\nu \Phi^J - \bar{V}(\Phi) \right],
    \label{SJordan}
\end{equation}
where $2f(\Phi)=M_{\rm Pl}^2 + \xi | \Phi|^2 $ with $\vert \Phi \vert^2 = \delta_{IJ} \Phi^I \Phi^J$ and $M_{\rm Pl} \equiv 1 / \sqrt{8\pi G} = 2.43 \cdot 10^{18} \, {\rm GeV}$ is the reduced Planck mass. The frame in which the nonminimal coupling $f (\Phi) \bar{R}$ remains explicit in the action is known as the ``Jordan frame''. The gravitational degrees of freedom can be written in the usual Einstein-Hilbert form upon performing a conformal transformation to the ``Einstein frame'',
\begin{equation}
    \bar{g}_{\mu \nu} =  \Omega^2 (x) \, g_{\mu \nu}, \ \text{with} \ \Omega^2(x)=1+ \xi \frac{| \Phi |^2  }{M_{\rm Pl}^2}.
    \label{conformaltrans}
\end{equation}
We can furthermore define the rescaled potential $V$ as
\begin{align}\label{eq:Vrescale}
    V(\Phi)=\frac{\bar{V}(\Phi)}{(1 + \xi |\Phi|^2/M_{\rm Pl}^2)^2} .
\end{align}
In terms of this potential and the transformed metric, the action in the Einstein frame then reads
\begin{align}
    S= \int d^4 x \sqrt{-g }\left[\frac{M_{\rm Pl}^2}{2} R - \frac{1}{2} \mathcal{G}_{IJ} g^{\mu\nu} \partial_\mu \Phi^I \partial_\nu \Phi^{J}- V(\Phi) \right],
    \label{SEinstein}
\end{align}
with the field-space metric \cite{Kaiser:2010ps}
\begin{align}
    \mathcal{G}_{IJ}=& \frac{M_{\rm Pl}^2}{M_{\rm Pl}^2 + \xi |\Phi|^2}\left[ \delta_{IJ}+ 6 \sigma \frac{\partial_I f \partial_J f}{M_{\rm Pl}^2 + \xi |\Phi|^2} \right] =  \frac{M_{\rm Pl}^2}{M_{\rm Pl}^2 + \xi |\Phi|^2}\left[ \delta_{IJ}+ 6 \sigma \xi^2 \frac{\delta_{IK} \delta_{JL} \Phi^K \Phi^L }{M_{\rm Pl}^2 + \xi |\Phi|^2} \right].
\label{eq:fsmetric}
\end{align}
In Eq.~(\ref{eq:fsmetric}), we have inserted the dimensionless parameter $\sigma$, which varies depending on one's assumption about the underlying gravitational theory.
The non-diagonal terms within ${\cal G}_{IJ}$ (proportional to $\sigma$) arise from the transformation of the connection $\Gamma^\mu_{\nu\lambda}$ under the conformal transformation of the spacetime metric $g_{\mu\nu}$ in Eq.~(\ref{conformaltrans}), upon assuming that the $\Gamma^\mu_{\nu\lambda}$ take the standard Levi-Civita form. On the other hand, in the Palatini (or metric-affine) formulation, the Christoffel symbols $\Gamma^\mu_{\nu\lambda}$ enter the action as independent degrees of freedom from $g_{\mu\nu}$ \cite{Bauer:2010jg}. (They are related, in this case, by the equations of motion.) In that case, the $\Gamma^\mu_{\nu\lambda}$ do not change under the conformal transformation of $g_{\mu\nu}$ in Eq.~(\ref{conformaltrans}).
We therefore set $\sigma = 1$ for standard metric gravity, and $\sigma = 0$ for the Palatini formulation.

\subsection{Single field models}\label{sec:EFTsinglefield}
For the case of a single scalar field $\Phi=\phi$, the field-space metric~\eqref{eq:fsmetric} reduces to a single function, ${\cal G}_{IJ} (\Phi) \rightarrow {\cal G}_{\phi\phi} (\phi)$. This allows for a field redefinition $\phi \to \chi$ to a canonically normalized field $\chi$.
For metric gravity ($\sigma=1$), this is defined through
\begin{equation}\label{eq:CanNon1}
    \frac{d\phi (\chi)}{d\chi}= \frac{\Omega^2}{( \Omega^2 +6 \xi^2 \phi^2/M_{\rm Pl}^2 )^{\frac{1}{2}}} .
\end{equation}
This differential equation has an analytical solution for $\chi (\phi)$. In terms of the auxiliary variable $\alpha \equiv 1+1/(6\xi)$ we find
\begin{align}\label{CanNon2}
    \chi(\phi) = & \sqrt{6}M_{\rm Pl} \tanh ^{-1}(\alpha^{1/2}) - \sqrt{6}M_{\rm Pl} \alpha^{1/2} \, \log \left(\frac{\sqrt{6 (\alpha^2+1)+\alpha \left(\varphi ^2-12\right)}-\alpha^{1/2} \varphi }{\sqrt{6} \left(\alpha-1\right)}\right) \nonumber  \\ 
    &\quad\>  -  \sqrt{6}M_{\rm Pl}\tanh ^{-1}\left(\alpha^{1/2} + \frac{\alpha^{1/2} \varphi^2-\varphi  \sqrt{6 (1 + \alpha^2) + \alpha \left(\varphi^2-12\right)}}{6 \left(\alpha-1\right)} \right),
\end{align}
where $\varphi \equiv \phi / M_{\rm Pl}$. Expanding around $\phi_c \sim 0$, and taking $\xi \gg 1/6$, this relation can be expanded as
\begin{align}
    \chi(\phi) = & \phi \left[ 1 + \frac{\xi^2 \phi^2}{M_{\rm Pl}^2} - \frac{9}{10} \frac{\xi^4 \phi^4}{M_{\rm Pl}^4} +\frac{27}{14} \frac{\xi^6 \phi^6}{M_{\rm Pl}^6} -\frac{45}{8} \frac{\xi^8 \phi^8}{M_{\rm Pl}^8} +\frac{1701}{88} \frac{\xi^{10} \phi^{10}}{M_{\rm Pl}^{10}}  +... \right] .
\end{align}
This series can then be inverted order by order in $\xi^2 \chi^2/M_{\rm}^2$,
\begin{align}\label{eq:phiofchiMetric}
    \phi(\chi)\simeq \chi \bigg[1 - \frac{\xi^2 \chi^2}{M_{\rm Pl}^2} + \frac{39}{10} \frac{\xi^4 \chi^4}{M_{\rm Pl}^4} - \frac{1479}{70} \frac{\xi^6 \chi^6}{M_{\rm Pl}^6} + \frac{37369}{280} \frac{\xi^8 \chi^8}{M_{\rm Pl}^8} - \frac{14196747}{15400} \frac{\xi^{10} \chi^{10}}{M_{\rm Pl}^{10}} + ... \bigg].
\end{align}
Considering for concreteness a Higgs-like potential in the Jordan frame at high energies,
\begin{equation}\label{eq:Vquartic}
    \bar{V}(\phi)= \frac{\lambda}{4} \phi^4,
\end{equation}
this allows us to rewrite the rescaled potential of Eq.~(\ref{eq:Vrescale}) in familiar EFT form. In the limit $\xi \gg 1/6$ the dominant terms arise from the redefinition of the $\phi^4$-term, while the conformal factor contributes higher-dimensional terms suppressed by $\xi/M_{\rm Pl}^2$:
\begin{align}\label{eq:VMetric}
    V(\chi) \equiv& \frac{\lambda}{4} \chi^4 + \sum_{n=1}^{\infty} C^{V,M}_{2n+4} \frac{ \xi^{2n} \chi^{2n+4} }{M_{\rm Pl}^{2n}} = \frac{\lambda}{4} \chi^4 - \lambda \frac{\xi^2 \chi^6 }{M_{\rm Pl}^2}+ \frac{27 \lambda }{5 } \frac{\xi^4 \chi^8}{M_{\rm Pl}^4} - \frac{1184 \lambda }{35} \frac{\xi^6 \chi^{10}}{M_{\rm Pl}^6} + ...
\end{align}
(We adopt the notation $C^{V,M}$ to indicate that these Wilson coefficients arise from terms in the potential within the metric-gravity formulation.) Crucially, we observe that all Wilson coefficients are proportional to $\lambda$. This implies, in particular, that they vanish for the case of a free scalar field, in agreement with Ref.~\cite{Hertzberg:2010dc}. 

\subsection{Multifield models}\label{sec:EFTmultifield}

The situation changes drastically if multiple scalar fields are involved. In particular, if two or more scalar fields have nonminimal couplings to the spacetime Ricci scalar, then in general the induced field-space manifold in the Einstein frame is not conformal to flat, and no combination of field redefinitions can make the kinetic terms for each scalar field canonical. In other words, there do not exist any field redefinitions akin to Eq.~(\ref{eq:CanNon1}) for which ${\cal G}_{IJ} \rightarrow \delta_{IJ}$. (For a detailed discussion, see Ref.~\cite{Kaiser:2010ps}.) We therefore describe the corresponding interactions by expanding the field-space metric of Eq.~(\ref{eq:fsmetric}) in the limit $\langle \phi \rangle \simeq 0$ as
\begin{equation}
\begin{split}
    \mathcal{G}_{IJ} &=  \delta_{IJ} +6 \frac{\xi^2}{M_{\rm Pl}^{2}} \delta_{IM} \delta_{NJ} \Phi^M \Phi^N +\delta_{IJ} \left[\sum_{n=1}^\infty (-1)^n \frac{\xi^{n} |\Phi|^{2n} }{M_{\rm Pl}^{2n}} \right]  \\ 
    &+6 \frac{\xi^2}{M_{\rm Pl}^2} \left[\sum_{n=1}^\infty (-1)^n(1+n) \frac{\xi^{n} |\Phi|^{2n} }{M_{\rm Pl}^{2n}} \right] \delta_{IM} \delta_{NJ} \Phi^M \Phi^N.
\end{split}
\label{eq:GIJexpandmetric}
\end{equation}
Similar to the case of a single field, this implies that for large nonminimal couplings $\xi$ the leading-order interaction is suppressed by $\xi/M_{\rm Pl}$, and linked to the non-diagonal terms in ${\cal G}_{IJ}$ that arise from the transformation of the connection. Meanwhile, the higher-dimensional terms arising within the re-scaled potential of Eq.~(\ref{eq:Vrescale}) are all suppressed by $\sqrt{\xi}/M_{\rm Pl}$,
\begin{align}\label{eq:Vexp}
    V(\Phi)=\frac{\lambda}{4}  |\Phi|^4 + \frac{\lambda}{4}  \sum_{n=1}^\infty (-1)^n (1+n)  \frac{\xi^n}{M_{\rm Pl}^{2n}} |\Phi|^{2n+4}.
\end{align}
If promoted to a gauge theory, the high-energy scattering among multiple scalars captures the scattering into the longitudinal modes of the gauge bosons by the Goldstone-boson equivalence theorem. We do not study couplings to fermions in detail here, but we have checked that for the models of interest, the corrections from Yukawa interactions are small for reasonable values of the Yukawa couplings. (Some details are provided in Eq.~(\ref{Afermion}) when expanding around the inflationary background.)

\subsection{Palatini gravity}\label{sec:EFTPalatini}

So far, we have found that the most important higher-dimensional terms in the effective action are linked to the transformation of the Levi-Civita connection. The transformation properties of the connection $\Gamma^\mu_{\nu\lambda} (x)$, however, are not unique. In Palatini (or metric-affine) gravity, for instance, the connection is treated as a dynamical degree of freedom, independent of the spacetime metric $g_{\mu\nu} (x)$ \cite{Bauer:2010jg}, and therefore $\Gamma^\mu_{\nu\lambda} (x)$ does not change under the transformation $\bar{g}_{\mu\nu} \rightarrow g_{\mu\nu}$ of Eq.~(\ref{conformaltrans}). In our context, this amounts to a different field-space metric after the conformal transformation, setting $\sigma = 0$ in Eq.~(\ref{eq:fsmetric}) and expanding around $\langle \phi \rangle \simeq 0$,
\begin{equation}\label{eq:fsmetricPal}
    \mathcal{G}_{IJ}= \frac{M_{\rm Pl}^2}{M_{\rm Pl}^2 + \xi |\Phi|^2} \delta_{IJ}= \left[\sum_{n=0}^\infty (-1)^n \frac{\xi^{n} |\Phi|^{2n} }{M_{\rm Pl}^{2n}} \right] \delta_{IJ}.
\end{equation}
For the potential in the Einstein frame, meanwhile, we recover Eq.~\eqref{eq:Vexp}.

In the case of a single scalar field, we can again canonically normalize the latter through a simplified version of Eq.~\eqref{eq:CanNon1},
\begin{equation}\label{eq:PCanNon1}
    \frac{d\phi (\chi)}{d\chi}= \Omega = \left( 1 + \xi \frac{\phi^2}{M_{\rm Pl}^2} \right)^{1/2} .
\end{equation}
This equation can be solved analytically for both $\phi(\chi)$ as well as $\chi(\phi)$,
\begin{align}
    \chi (\phi)=& \frac{M_{\rm Pl}}{\sqrt{\xi }}\sinh ^{-1}\left(\frac{\sqrt{\xi} \phi }{M_{\rm Pl}} \right),  \\ 
    \phi (\chi) = & \frac{M_{\rm Pl}}{\sqrt{\xi}} \sinh\left( \frac{\sqrt{\xi} \chi }{M_{\rm Pl}}\right) . \label{eq:phiofchiPalatini}
\end{align}
The higher-dimensional terms are now indeed suppressed by $\sqrt{\xi}/M_{\rm Pl}$ (as well as numerically small coefficients) rather than $\xi/M_{\rm Pl}$. This suggests the higher-dimensional terms in the potential arising from the rescaling by the conformal factor to be relevant, unlike in our discussion in Sec.~\ref{sec:EFTsinglefield}. Using Eq.~\eqref{eq:phiofchiPalatini}, it is straightforward to obtain the potential in terms of the canonically normalized field $\chi$. Again considering the quartic potential of Eq.~(\ref{eq:Vquartic}) for concreteness and expanding around $\langle \phi \rangle \simeq 0$, we find
\begin{align}\label{eq:VPalatini}
    V(\chi) = & \frac{\lambda}{4} \frac{M_{\rm Pl}^4}{\xi^2}\tanh^4 \left( \frac{\sqrt{\xi} \chi}{M_{\rm Pl}} \right) = \frac{\lambda}{4} \chi^4 - \frac{\lambda}{3 } \frac{\xi  \chi^6}{M_{\rm Pl}^2} + \frac{3\lambda }{10} \frac{\xi \chi^8}{M_{\rm Pl}^4} - \frac{212 \lambda }{945} \frac{\xi^3 \chi^8}{M_{\rm Pl}^6} + ...
\end{align}
Comparing the higher-dimensional operators of this potential with the ones in the effective potential obtained from metric gravity in Eq.~\eqref{eq:VMetric}, we observe that the higher-dimensional terms are not only suppressed by a larger scale $M_{\rm Pl}/\sqrt{\xi}$, but also through notably smaller numerical coefficients. 

\section{Precise unitarity bounds}\label{sec:UnitarityBounds}

We are now ready to apply the unitarity analysis developed in Ref.~\cite{Abu-Ajamieh:2020yqi} (also reviewed in Ref.~\cite{Cohen:2021ucp}) to the effective theory representing a nonminimal coupling. For the sake of self-consistency we first repeat the derivation of the main equations provided by these authors before applying them to the different cases and their corresponding towers of operators discussed in the previous section. As described above, in this section we expand the effective action around the configuration $\langle \phi \rangle \simeq 0$.

\subsection{$S$-matrix formulation}
We set out from the $S$-matrix, which we expand as
\begin{equation}
    S= 1 - i T.
\end{equation}
For our analysis, we will consider the scattering of $M=\{m_1, ... , m_W\}$ ingoing scalar particles into $N=\{n_1,...,n_W\}$ outgoing scalar particles, where $n_I$ and $m_I$ respectively denote the number of particles of the species $I$. Following Ref.~\cite{Cohen:2021ucp}, we denote their corresponding states $|Q,M \rangle$ and $|P,N\rangle$ respectively, where $Q$ and $P$ denote the total momentum of these particles. Total momentum conservation can be made manifest through the normalization condition
\begin{equation}
    \langle P, M | Q, N \rangle = (2 \pi)^4 \delta^{(4)}(P-Q) \delta_{M N} .
\end{equation}
We can describe this process through the corresponding matrix elements of $S$ and $T$, $\mathcal{S}_{MN}$ and $\mathcal{M}_{MN}$, defined by
\begin{align}
    \mathcal{S}_{MN} \cdot (2 \pi)^4 \delta^{(4)}(P-Q)  =& \langle P, M | S | Q, N \rangle , \\
    \mathcal{M}_{MN} \cdot (2 \pi)^4 \delta^{(4)}(P-Q)  =& \langle P, M | T | Q, N \rangle .
\end{align}
Using the optical theorem, we can now represent the unitarity of the $S$-matrix during this process in terms of its matrix elements 
\begin{equation}
    1=\sum_N |\mathcal{S}_{MN}|^2 = 1 + 2 {\rm Im}(\mathcal{M}_{MM}) +\sum_N |\mathcal{M}_{M N}|^2.
\end{equation}
This implies in particular separate upper bounds on the matrix elements representing elastic and inelastic scatterings,
\begin{gather}\label{eq:boundsum}
    \sum_{M \neq N} |\mathcal{M}_{MN}|^2 = \sum_{N \neq 2} |\mathcal{S}_{MN}|^2  \leq 1.
\end{gather}
The matrix elements $\mathcal{M}_{MN}$ are related to the scattering amplitudes $|Q,M\rangle \to |P, N \rangle$, denoted by $\mathcal{A}_{M \to N}$, through
\begin{align}\label{eq:qab}
    \mathcal{M}_{MN}= &\left[m_1! ... n_S! \cdot{\rm Vol}_M \cdot {\rm Vol}_N  \right]^{- \frac{1}{2}}  \intd \rho_M \intd \rho_N \  \mathcal{A}_{M \to N},
\end{align}
where ${\rm Vol}_N$ denotes the Lorentz-invariant $N$-particle phase space volume. The latter is defined through its corresponding measure,
\begin{equation}\label{eq:drho}
    {\rm d}\rho_N= (2 \pi)^4 \delta^{(4)}\left( K - \sum_{i=1}^N k_i \right) \prod_{i=1}^N \left( \frac{{\rm d}^3 k_i}{2 (2\pi)^3 |\vec{k}_i|} \right).
\end{equation}
The total volume of its corresponding phase space has been shown to take the simple form~\cite{Abu-Ajamieh:2020yqi}
\begin{equation}\label{eq:LIPSVol}
    {\rm Vol}_N \equiv \int {\rm d}\rho_N = \frac{1}{8 \pi (N-1)! (N-2)!} \left( \frac{E}{4 \pi} \right)^{2(N-2)}\hspace{-0.5cm},
\end{equation}
where the center-of-mass energy $E$ is defined as $E^2=Q^2=P^2$. Moreover, we will require the integrals~\cite{Abu-Ajamieh:2020yqi}
\begin{align}
    \int k_1^\mu  {\rm d}\rho_N =& \frac{K^\mu}{N} {\rm Vol}_N \label{eq:LIPSp} ,\\ 
    \int k_1^\mu (k_2)_\mu {\rm d}\rho_N=& \frac{(N-2)!}{N!} E^2 {\rm Vol}_N .
    \label{eq:LIPSpp} 
\end{align}

\subsection{Single-field model}
Following our discussion in Sec.~\ref{sec:EFTsinglefield}, in the case of a single field the nonminimal coupling can be absorbed entirely into higher-dimensional operators in the potential through the field redefinitions of Eqs.~\eqref{eq:phiofchiMetric} and~\eqref{eq:phiofchiPalatini}, respectively. 

For simplicity and to allow for easier comparison with previous works~\cite{Burgess:2009ea,Burgess:2010zq,Barbon:2009ya,Giudice:2010ka,Lerner:2011it,Barbon:2015fla}, we will consider the scattering of two scalar particles into arbitrary initial states. At tree-level, the matrix elements arising from the higher-dimensional operators are then given by
\begin{align}
    \mathcal{A}_{2 \to N}= (N+2)! \cdot \frac{C_{N+2}}{\Lambda^{N-2}},\ \ \text{for} \ \ N\in 4 +2 \mathbb{N}.
\end{align}
In metric gravity, we have $\Lambda^2= M_{\rm Pl}^2/\xi^2$, whereas in Palatini gravity $\Lambda^2= M_{\rm Pl}^2/\xi$. This expression is momentum-independent, allowing one to replace the phase space integrals in Eq.~\eqref{eq:qab} with the corresponding volumes~\eqref{eq:LIPSVol}. The corresponding matrix element therefore reduces to
\begin{align}
    \mathcal{M}_{2N}= & \left(\frac{{\rm Vol}_2 {\rm Vol}_N}{2\cdot N!}\right)^{1/2}  (N+2)! \cdot \frac{C_{N+2}}{\Lambda^{N-2}} = \nonumber \\ 
    =&\frac{1}{8 \pi} \frac{(N+2)!}{\left[2 \cdot N! (N-1)! (N-2)! \right]^{1/2}} C_{N+2} \left( \frac{E}{4\pi \Lambda} \right)^{N-2} . \label{eq:M2N}
\end{align}
To understand the significance of the prefactors arising from the phase space volume and symmetry factors we first consider the limit of large $N$, where the matrix elements asymptote to
\begin{align}
    \mathcal{M}_{2N}\to \frac{C_{N+2}}{(N!)^{1/2}}\left( \frac{E}{4\pi \Lambda}\right)^N \simeq \left( \frac{\sqrt{e}}{\sqrt{N}} C_{N+2}^{1/N} \frac{E}{4\pi \Lambda}\right)^N,
\end{align}
where in the second step we have used Stirling's formula. Demanding $|M_{2,N}|^2\leq 1$ for each term in Eq.~\eqref{eq:boundsum} then implies an individual upper bound on $E$ for each $N$,
\begin{align}
    E< 4\pi \Lambda \frac{\sqrt{N}}{\sqrt{e} C_{N+2}^{1/N}}.
\end{align}
The $N$-dependence of the right-hand side now depends on the scaling of the Wilson coefficients $C_N$. If they grow \textit{faster} than $C_{N+2} \sim  \sqrt{N/e}^N \sim\sqrt{N!}$, we find that larger $N$s correspond to increasingly stronger bounds on $E$, implying severe unitarity issues. In the cases of interest here, however, we find a \textit{slower} growth of $C_N$. Thus, the additional factor of $\sqrt{N}$ implies that larger $N$ give rise to weaker bounds.

This simple scaling relation also leads to a dependence of the unitarity bound on the numerical value of the coupling: Without this loosening of the bound for large $N$, the bound would be dominated by contributions from large $N$, for which $C_{N+2}^{1/N}\to 1$ irrespective of the strength of the coupling. Considering the full Eq.~\eqref{eq:M2N} on the other hand implies that the most stringent bound arises from relatively \textit{small} $N$, for which the dependence on the Wilson coefficients remains relevant. 

\begin{figure}[t!]
    \centering
    \includegraphics[width=0.98 \textwidth]{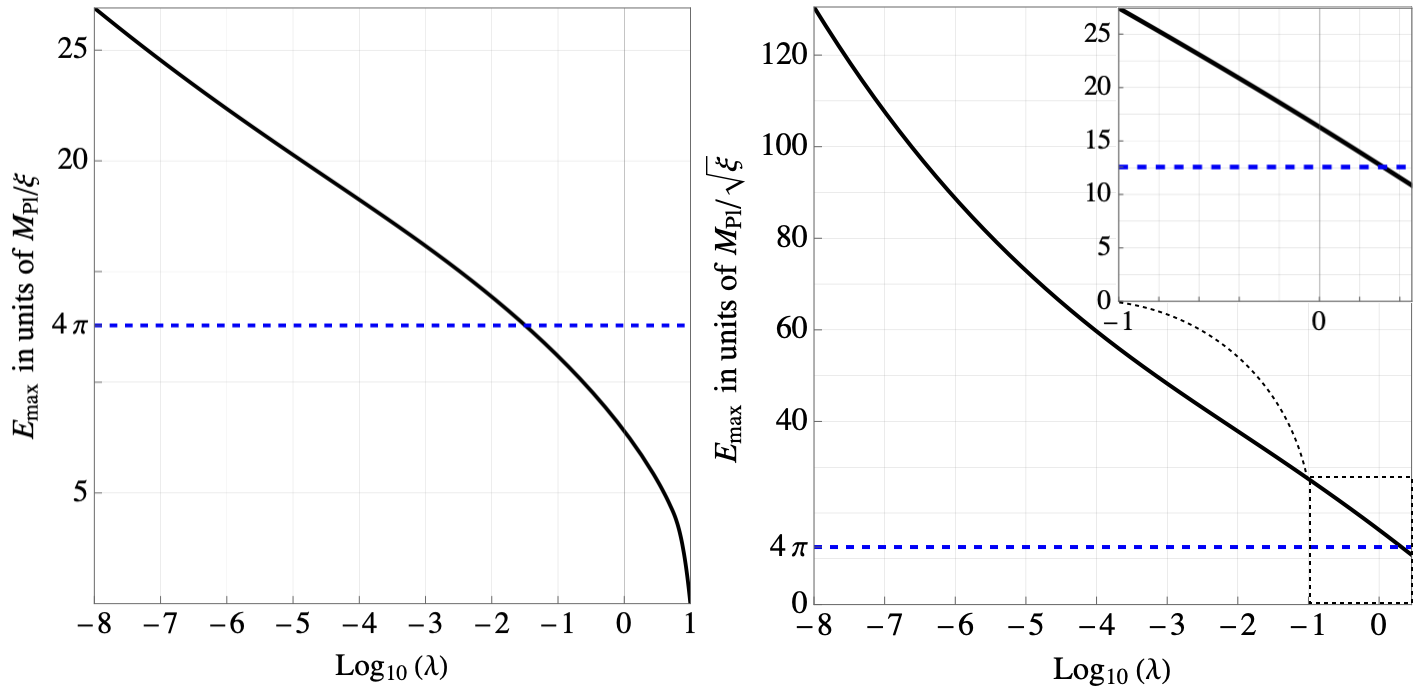}
    \caption{\textit{Left:} The upper bound on the energy $E_{\rm max}$ obtained by demanding unitarity arising from the Jordan frame potential $\bar{V}=\lambda \phi^4 /4$ in metric gravity as a function of $\lambda$. \textit{Right:} The same bound assuming Palatini gravity.}
    \label{fig:BoundsSingleField}
\end{figure}

We first derive the precise unitarity bound for a single scalar field in metric gravity, which is determined by Eq.~\eqref{eq:boundsum}. Using Eqs.~\eqref{eq:boundsum} and~\eqref{eq:M2N}, this translates to
\begin{align}\label{EmetricV}
    \sum_{N\in 2+2 \mathbb{N}}  \frac{1}{128 \pi^2} \frac{\left[(N+2)! \cdot C_{N+2}^{V,M}\right]^2}{ N! (N-1)! (N-2)!} \left(\frac{E \cdot \xi}{4 \pi M_{\rm Pl}}\right)^{2(N-2)}  \leq 1.
\end{align}
Note that here, as in the following, we have taken into account the contribution of the elastic scattering process $2\phi \to 2\phi$. For the potential~$\bar{V}= \lambda \phi^4/4$, this inequality can be understood as an upper bound on the energy $E$ as a function of $\lambda$. We present our results for this bound in the left panel of Fig.~\ref{fig:BoundsSingleField}. In agreement with our previous argument we find a notable dependence of the upper bound on $\lambda$. In particular, for the range of small couplings favored, e.g., by scenarios such as critical Higgs inflation, $\lambda \sim 10^{-5}$, unitarity is restored up to
$E_{\rm max} \sim \mathcal{O}(20) \,M_{\rm Pl}/\xi$. This represents a significant deviation from the naive estimate obtained through power-counting alone, even after we perform the sum over all outgoing states. Moreover, this establishes a smooth transition to the limit of a free scalar field, for which it was shown that the nonminimal coupling does not cause any unitarity issues~\cite{Hertzberg:2010dc}. Lastly, we observe that in this regime it would indeed be justified to neglect the inelastic scattering in Eq.~\eqref{eq:boundsum}, whose corresponding amplitude scales simply by $\lambda$.

The case of Palatini gravity can be investigated by the same procedure, where the result takes the form
\begin{align}
\sum_{N\in 2+2 \mathbb{N}} \frac{1}{128 \pi^2}  \frac{\left[ (N+2)! \cdot C_{N+2}^{V,P} \right]^2}{ N! (N-1)! (N-2)!} \left(\frac{E \cdot \sqrt{\xi}}{4 \pi M_{\rm Pl}}\right)^{2(N-2)} \leq 1,
\end{align}
with the Wilson coefficients $C^{V,P}$ given in Eq.~\eqref{eq:VPalatini}. We present this bound in the right panel of Fig.~\ref{fig:BoundsSingleField}. In agreement with our earlier observation of numerically small Wilson coefficients we find that, for any given $\lambda$, the scale of unitarity violation is further increased relative to that metric gravity.

\subsection{Multifield models}
For models with multiple scalar fields, we had found that introducing a nonminimal coupling can be understood as the introduction of new terms to the field-space metric. Unlike for the case of a single field, these \textit{cannot} be absorbed into the potential, and hence need to be treated directly as momentum-dependent interactions. 

\subsubsection{Metric gravity}

In metric gravity, we had found in Eq.~(\ref{eq:GIJexpandmetric}) that when expanding the action around the configuration $\langle \phi \rangle \simeq 0$, there exists a single higher-dimensional operator suppressed by $\xi/M_{\rm Pl}$,
\begin{equation}\label{eq:momdepint}
    \frac{-\Delta {\cal L}}{\sqrt{-g}} = 3 \frac{\xi^2}{M_{\rm Pl}^2} g^{\mu\nu} \delta_{IM} \delta_{NJ} \Phi^M \Phi^N \partial_\mu \Phi^I \partial_\nu \Phi^J ,
\end{equation}
whereas all other higher-dimensional operators are suppressed by the parametrically larger scale $M_{\rm Pl}/\sqrt{\xi}$. This suggests that the sum in Eq.~(\ref{eq:momdepint}) is dominated by the elastic scattering process $2 \phi \to 2 \phi$. In the following we consider the scattering process $\phi_1+\phi_2 \to \phi_1+\phi_2 $ for concreteness (and also to allow for an easier comparison with the existing literature~\cite{Hertzberg:2010dc}). Considering, on the other hand, scattering processes $2 \phi^I \to 2 \phi^J$ would give rise to an independent value of $E_{\rm max}$. 
Unitarity requires $\Lambda_{\rm NP}$ to be  smaller than both of these values. In the limit of a large number of species $W$, the latter bound can, in fact, be expected to be dominant due to larger number of possible outgoing states, corresponding to a scaling $E_{\rm max} \propto W^{-1/2}$. However, for $W=4$, which is our primary focus, we expect these bounds to be within an $\mathcal{O}(1)$ factor of each other.

For the corresponding amplitude, we find 
\begin{align}
    \mathcal{A}_{2\to 2}= \frac{12 \xi^2}{M_{\rm Pl}^2}\left( (q_1)_\mu  p_1^\mu + (q_2)_\mu p_2^\mu \right)\to \frac{24 \xi^2}{M_{\rm Pl}^2} (q_1)_\mu \cdot p_1^\mu.
\end{align}
Here, $q_1$ and $q_2$ denote the momenta of the ingoing particles and $p_1$ and $p_2$ those of the outgoing ones. In the second step, we have used that both of the terms will lead to identical contributions once we integrate over the in- and outgoing momenta. Setting $N=2$ in Eq.~\eqref{eq:qab}, we find for the relevant matrix element
\begin{align}
    \mathcal{M}_{22}= 8 \pi \cdot \frac{24 \xi^2}{M_{\rm Pl}^2} \int {\rm d}\rho_2(q_1) (q_1)_\mu  \int {\rm d}\rho_2(q_1) (p_1)^\mu .
\end{align}
Using the integral identity Eq.~\eqref{eq:LIPSp}, it is straightforward to see the only non-vanishing components of the integrals on the right-hand side are the temporal components $\mu=0$, for which we find
\begin{align}
     \int {\rm d}\rho_2(q_1) (q_1)_0 = \frac{E}{16 \pi}.
\end{align}
Altogether, this implies that
\begin{align}
    M_{22}=\frac{3}{4 \pi} \left( \frac{E\cdot \xi}{M_{\rm Pl}}\right)^2 .
\end{align}
Thus we find in agreement with earlier works that the term~\eqref{eq:momdepint} leads to a unitarity violation bound that is independent of the coupling $\lambda$,
\begin{equation}
   E_{\rm max} = \sqrt{\frac{4 \pi}{3}} \frac{M_{\rm Pl}} {\xi} \simeq 2.05 \frac{ M_{\rm Pl}}{\xi}.
\label{Cutmetric}
\end{equation}
Comparing with the cut-off scale arising from terms in the potential, Eq.~(\ref{EmetricV}), we see that in the multifield case the cut-off scale in Eq.~(\ref{Cutmetric}), arising from the momentum-dependent terms in the noncanonical kinetic sector, places a more stringent unitarity requirement, with $E_{\rm max} \sim M_{\rm Pl} / \xi$ rather than ${\cal O}(10) \cdot M_{\rm Pl} / \xi$. 

\subsubsection{Palatini gravity}\label{sec:UnitarityPalatini}
In Palatini gravity, the interactions arising from the field space metric in Eq.~\eqref{eq:fsmetric} and the potential are both suppressed by powers of $\sqrt{\xi}/M_{\rm Pl}$. Assuming a relatively small coupling constant then implies an additional suppression of the contributions from the latter to the $S$-matrix elements, allowing us to focus on the interaction terms
\begin{equation}
 \frac{-\Delta {\cal L}}{\sqrt{-g}} = \frac{1}{2} \delta_{IJ} \left[ \sum_{n = 1}^\infty (-1)^n \frac{ \xi^n \vert \Phi \vert^{2n}}{M_{\rm Pl}^{2n}} \right] g^{\mu\nu} \partial_\mu \Phi^I \partial_\nu \Phi^J .
    \label{eq:DeltaLPalatini}
\end{equation}
To make more explicit the structure of the additional vertices, we expand the coefficient $\vert \Phi \vert^{2n}$ using the multinomial theorem,
\begin{align}\label{eq:DeltaLPalatini2}
    \vert \Phi \vert^{2n}= \sum_{k_1+...+k_4=n} \frac{n!}{k_1! \cdot ... \cdot k_4!}(\phi^1)^{2k_1} \cdot ... \cdot (\phi^W)^{2k_4}.
\end{align}
We can now consider the scattering process $\phi_1 + \phi_2 \to n_1 \phi_1 + ... n_4 \phi_4$, with $N\equiv n_1+...+n_4$. Using Eqs.~\eqref{eq:boundsum} and~\eqref{eq:drho}, it is straightforward to obtain the corresponding matrix element as
\begin{align}\label{eq:MPal}
    \mathcal{M}=&  \frac{\pi}{2} \left(\frac{n_1! ... n_4!}{(N-2)!(N-1)!}\right)^{\frac{1}{2}} \frac{(3N-4) \left( \frac{N}{2} \right)! (n_1+1)(n_2+1) }{N (N-1) \left( \frac{n_1+1}{2} \right)!\left( \frac{n_2+1}{2} \right)!\left( \frac{n_3}{2} \right)!\left( \frac{n_4}{2} \right)!}\left( \frac{\sqrt{\xi} E}{4 \pi M_{\rm Pl}} \right)^{N}.
\end{align}
For the details of this derivation, see appendix~\ref{sec:MatrixPalatini}. Similar to the single field case, we find that the volume factors cause these elements to vanish for large numbers of out going particles. Taking, as an example $n_1=N-1,n_2=1,n_3=n_4=0$ with $N\gg 1$, we find that $\mathcal{M}\sim 1/\sqrt{N!}$ in the limit $N\gg 1$. 

Using Eq.~\eqref{eq:boundsum}, it is straightforward to numerically obtain an upper bound on the center-of-mass-energy $E$,
\begin{align}
    E_{\rm max} \simeq 4.94 \frac{M_{\rm Pl}}{\sqrt{\xi}}.
\label{CutPalatini}\end{align}
Similar to the case of metric gravity in Eq.~(\ref{Cutmetric}), this only represents a minor improvement over the common estimate, albeit by a relatively larger numerical factor.

\subsubsection{Canonical kinetic model}

\begin{figure}[t!]
    \centering
    \includegraphics[width=0.75 \textwidth]{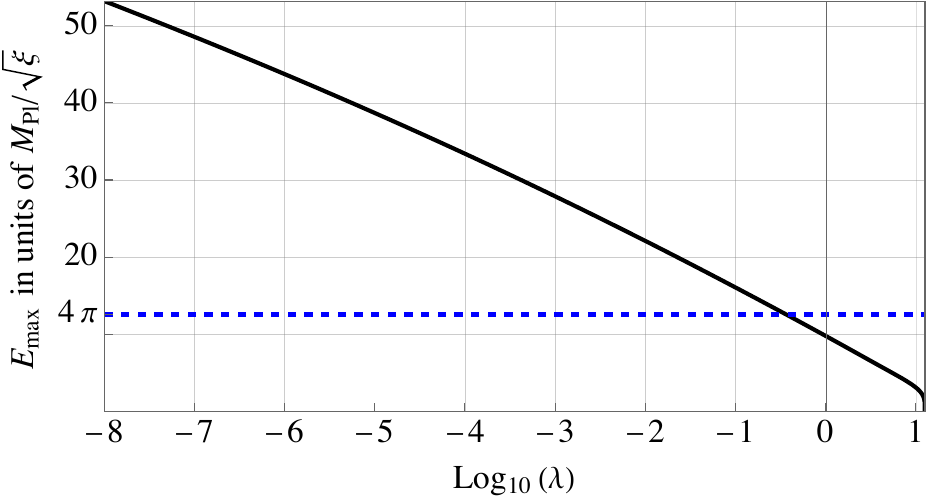}
    \caption{The upper bound on the energy obtained by demanding unitarity arising from the Lagrangian~(\ref{Lcan}) as a function of $\lambda$.}
    \label{fig:BoundsMarksLagrangianFull}
\end{figure}

In both single and multifield models we have found that the higher-dimensional operators contributing to the unitarity violation arise from two separate sources: The rescaling of the potential by the conformal factor $\Omega$ and the nontrivial field space metric $\mathcal{G}_{IJ}$. Having analyzed the contributions to the $S$-matrix arising from the latter, we can now consider separately the interactions linked to the rescaling of the potential. This amounts to considering a toy-model Lagrangian
\begin{align}
    \frac{\mathcal{L}}{\sqrt{-g}} = -\frac{1}{2}g^{\mu \nu} \delta_{IJ}\partial_\mu \Phi^I \partial_\nu \Phi^J - \frac{\lambda}{4}\cdot \frac{|\Phi|^4}{\left(1+\frac{\xi |\Phi|^2}{M_{\rm Pl}^2}\right)^2}.
\label{Lcan}\end{align}
To expand the potential in terms of the individual fields $\Phi^I$, we can use the identity
\begin{align}
    \frac{1}{(1+x)^2}=\sum_{n=0}^\infty (-1)^n (1+n) x^n.
\end{align}
Together with the multinomial theorem~\eqref{eq:DeltaLPalatini2}, this implies
\begin{align}
    V(|\Phi|)= \frac{\lambda}{4 \Lambda^{2n}} \sum_{n=0}^\infty  \sum_{k_1+...+k_4=n+2}  \frac{(-1)^n (1+n)  (n+2)!}{k_1! \cdot ... \cdot k_4!}\phi_1^{2k_1} \cdot ... \cdot \phi_4^{2k_4}. 
\end{align}
To relate the corresponding $S$-matrix to our results in the previous subsections, we again consider the scattering process of one $\phi_1$ particle with one $\phi_2$ particle. Again labeling the outgoing states by the number of outgoing particles of each species by $n_i$, it is straightforward to find
\begin{align}
    \mathcal{A}= \frac{\lambda}{8 \Lambda^{N-2}}  \frac{ N \left(\frac{N}{2}+1\right)!}{\left(\frac{n_1-1}{2}\right)! \left(\frac{n_2-1}{2}\right)! \left(\frac{n_3}{2}\right)! \left(\frac{n_4}{2}\right)!} \cdot (n_1+1)!(n_2+1)! n_3! n_4!.
\end{align}
For the total matrix element, this implies
\begin{align}
    \mathcal{M}=&\frac{\lambda}{256 \pi}  \frac{\left(\frac{N}{2} + 1\right)! N (n_1+1)^2 (n_2+1)^2}{\left(\frac{n_1+1}{2}\right)!\left(\frac{n_2+1}{2}\right)!\left(\frac{n_3}{2}\right)!\left(\frac{n_4}{2}\right)!} \left[  \frac{n_1! n_2! n_3! n_4!}{(N-1)!(N-2)!}  \right]^{1/2} ,
\end{align}
where $N=n_1+...+n_4$ as before. From here it is straightforward to obtain the upper bound on $E$ by summing $|\mathcal{M}|^2$ over all outgoing states and demanding the sum to be smaller than one. We present the corresponding bound on energy as a function of $\lambda$ in Fig.~\ref{fig:BoundsMarksLagrangianFull}.
As an example, for $\lambda=10^{-5}$, the bound is
\begin{equation}
    E_{\rm max} \simeq  38{ M_{\rm Pl}\over\sqrt{\xi}}\,\,\,\,\,\text{for}\,\,\,\,\lambda=10^{-5}.
\label{CutCanonical}\end{equation}
Note that this cut-off scale is parametrically larger than the values associated with multifield models in either the metric or Palatini formulations, Eqs.~(\ref{Cutmetric}) and (\ref{CutPalatini}).

\section{Inflationary observables}\label{sec:InflationaryObservables}

In this section, we briefly consider the connection of the above models to cosmological observables.

In the Einstein frame, the potential of all models studied in this work is
\begin{equation}
    V(\Phi)=\frac{\lambda}{4}\frac{|\Phi|^4}{(1+\xi|\Phi|^2/M_{\rm Pl}^2)^2}.
\end{equation}
For the purpose of inflation, the field will roll radially in the $\Phi$-space, as angular motion will tend to be quickly damped out by Hubble friction \cite{Greenwood:2012aj}. This generically suppresses typical multifield behavior during inflation, such as turns within the field space,  amplification of isocurvature perturbations, and non-Gaussianity, each of which remain consistent with the usual single field predictions \cite{Kaiser:2012ak,Schutz:2013fua,Kaiser:2015usz}. To analyze the dynamics during inflation, we may therefore take 
$\Phi=\phi$, the single radial mode. Then different models are distinguished by their kinetic terms, as discussed in detail in Sec.~\ref{sec:EFT}.

\subsection{Metric gravity}
\label{MetricObs}

In the standard metric theory, the kinetic term in the Einstein frame is
\begin{equation}
\frac{-\mathcal{L}_{\rm kin}}{\sqrt{-g}}=
\frac{1}{2}\frac{1}{(1+\xi\phi^2/M_{\rm Pl}^2)}\left(1+6\frac{\xi^2\phi^2/M_{\rm Pl}^2}{1+\xi\phi^2/M_{\rm Pl}^2}  \right) g^{\mu \nu}  \partial_\mu \phi \partial_\nu \phi .
\end{equation}
By transforming to a canonically normalized field $\chi$ and then exploring the inflationary regime, which occurs at large $\chi$, the potential becomes
\begin{equation}\label{Vchimetric}
    V(\chi)\approx\frac{\lambda M_{\rm Pl}^4}{4\xi^2}\left(1-2
   e^{-\sqrt{2/3}\,\chi/M_{\rm Pl}} \right).
\end{equation}
Within the potential slow-roll approximation (PSRA) \cite{Liddle:1994dx}, we may identify the slow roll parameters $\epsilon$ and $\eta$ as
%
\begin{align}
&    \epsilon=\frac{M_{\rm Pl}^2}{2}\left(\frac{V'}{V}\right)^2 ,\\
 &   \eta=M_{\rm Pl}^2\frac{V''}{V}.
    \end{align}
In the slow-roll regime these are approximated by
\begin{align}
    & \epsilon(\chi) \simeq \frac{4}{3}e^{-2\sqrt{2/3}\,\chi/M_{\rm Pl}} ,\\
       & \eta(\chi) \simeq -\frac{4}{3}e^{-\sqrt{2/3}\,\chi/M_{\rm Pl}} .
\end{align}
The end of inflation is defined by $\epsilon=1$. In this leading-order approximation for the potential, the corresponding field value is
\begin{equation}\label{chiendmetric}
    \chi_{\rm end} \simeq 
      \sqrt{3\over2}\,M_{\rm Pl}\,\ln(2(3+\sqrt{3})/3) ,
\end{equation}
which is independent of $\xi$ in this $\xi \gg 1$ limit. The leading-order estimate for $V (\chi)$ in Eq.~(\ref{Vchimetric}) is not precise towards the end of inflation. Nevertheless, what matters is only that the value $\chi_{\rm end}$ is small compared to the value during inflation.

The number of e-foldings of inflation from when the field has value $\chi$ to the end of inflation is given by
\begin{equation}\label{Neintegral}
    N_e(\chi)=\frac{1}{M_{\rm Pl}}\int_{\chi_{\rm end}}^\chi{d\chi'\over\sqrt{2\epsilon(\chi')}}.
\end{equation}
Carrying out this integral and then inverting to obtain the field value in terms of the number of e-foldings gives
\begin{equation}\label{eq:chiofNMetric}
\chi(N_e)=\sqrt{3\over2}\,M_{\rm Pl}\,\ln\left({4\over3}N_e\right),
\end{equation}
where we have taken the $N_e\gg1$ limit (which means the value of $\chi_{\rm end}$ is negligible).

There can be corrections to Eq.~(\ref{eq:chiofNMetric}). In the scenario in which the self-coupling becomes very small, it can lie close to a near-critical point for delicately chosen parameters. In this case the true potential can have a correction that deviates from the above estimate. If there is a near-critical point, then there can be even more e-foldings of inflation. For the purpose of this section, we shall assume that the potential at the CMB pivot scale is well approximated by Eq.~(\ref{Vchimetric}), but leave open the possibility that some near-critical behavior of the potential at field values smaller than those relevant for CMB-scale dynamics could yield a larger total number of e-folds of inflation than the estimate in Eq.~(\ref{Neintegral}). In such cases, the total number of e-foldings is actually
\begin{equation}
    N_{\rm e,tot}\geq N_e.
\end{equation}
When there is no critical or near-critical point then $N_{\rm e,tot}= N_e$, otherwise we can have $N_{\rm e,tot}> N_e$. Since a typical reference value is $N_{\rm e,tot}=55$, then in this case we can have $N_e<55$. 

The two slow-roll parameters can then be expressed in terms of the number of e-foldings as
\begin{align}\label{epsilonetaSR}
    &\epsilon(N_e)={3\over 4N_e^2},\\
    &\eta(N_e)=-{1\over N_e}.
\end{align}
The prediction for the spectral index and tensor-to-scalar ratio is (see, e.g., Ref.~\cite{Planck:2013jfk})
\begin{align}
    &n_s=1+2\eta-6\epsilon ,\\
    &r=16\epsilon,
\end{align}
which should be evaluated when the modes exiting the horizon eventually re-enter in the CMB era. This is typically in the vicinity of $N_e \simeq 55$ e-folds before the end of inflation. Evaluating this using Eqs.~(\ref{epsilonetaSR}) yields
\begin{align}
&    n_s=1-{2\over N_e}\approx0.964 , \\
&r_s={12\over N_e^2}\approx 0.004,
\end{align}
where in the last step we evaluated these for $N_e=55$. These values are consistent with precision measurements of the CMB: $n_s = 0.9649 \pm 0.0042$ ($68\%$ confidence interval) \cite{Planck:2018jri} and $r_s < 0.036$ \cite{BICEP:2021xfz}.

We also need to consider the variance of density perturbations. Within the slow-roll approximation, this is given by \cite{Planck:2013jfk}
\begin{equation}
    \Delta_s={V\over 24\pi^2M_{\rm Pl}^4\,\epsilon}.
\end{equation}
In the slow-roll inflationary regime, this evaluates to
\begin{equation}
    \Delta_s={N_e^2\lambda\over72\pi^2\xi^2}.
\end{equation}
We need to equate this to the observed value of density perturbations on large scales, namely \cite{Planck:2018jri}
\begin{equation}
    \Delta_{s, \rm obs}\simeq 2.1\cdot 10^{-9}.
\end{equation}
In order for this to match, the nonminimal coupling $\xi$ must take on a particular value, namely
\begin{equation}
 \xi\simeq 819 N_e\sqrt{\lambda}.
\label{xivalue}\end{equation}
As a useful reference value, $\lambda_5\equiv10^{-5}$ and taking $N_e=55$, this gives
\begin{equation}
\xi\simeq 142\sqrt{\lambda/\lambda_5},
\end{equation}
with corresponding field value at $N_e = 55$ e-foldings before the end of inflation of
\begin{equation}\label{chivaluemetric}
    \chi\approx 5.3\, M_{\rm Pl}.
\end{equation}
Then using the equation for the unitary cut-off scale $E_{\rm max} \simeq 2 M_{\rm Pl}/\xi$ that we found for this model in Eq.~(\ref{Cutmetric}), we find
\begin{equation}\label{Emaxlambda5}
    E_{\rm max}\approx 0.014\,M_{\rm Pl}
    \sqrt{\lambda_5/\lambda}.
\end{equation}
We note that the field displacement $\Delta\chi\approx 5.3 M_{\rm Pl}$ from the vacuum to the value during inflation is significantly larger than this cut-off scale. (Note we can use the canonically normalized field $\chi$ to determine this, but it is also valid for any field redefinition when the corresponding metric on field space is used.) So one may question the reliability of this effective field theory.

From another perspective, one may employ a more direct comparison between physical energy scales, as follows: At the end of inflation, the energy density is
\begin{equation}
    \rho({t_{\rm end}})\approx V(\phi_{\rm end}\approx M_{\rm Pl}/\sqrt{\xi})={\lambda M_{\rm Pl}^4\over 16\,\xi^2}.
    \label{rhoend}
\end{equation}
As the universe evolves, we anticipate fairly rapid thermalization during reheating, with time evolution
\begin{equation}
    \rho(t_{\rm reh})\approx \rho({t_{\rm end}}) e^{-3N_{\rm reh}},
    \label{rhoreh}
\end{equation}
where $N_{\rm reh}$ is the number of e-foldings of the reheating phase \cite{Amin:2014eta,Allahverdi:2020bys}. We expect $N_{\rm reh}\sim {\cal O} (1)$: reheating should be rather quick, since the Higgs field has significant couplings to the rest of the Standard Model. (Such behavior is indeed borne out in numerical simulations~\cite{Bezrukov:2008ut,Garcia-Bellido:2008ycs,Ema:2016dny,Sfakianakis:2018lzf,Nguyen:2019kbm,vandeVis:2020qcp,Bettoni:2021zhq,Laverda:2023uqv,Bettoni:2024ixe,Figueroa:2024asq}.) The corresponding temperature (using $\rho=g_*\pi^2 T^4/30$) is
\begin{equation}
    T_{\rm reh}={15^{1/4}\lambda^{1/4}e^{-3 N_{\rm reh}/4}\over 2^{3/4}\sqrt{\pi}\,g_*^{1/4}} \, \frac{ M_{\rm Pl}}{\sqrt{\xi}} .
    \label{Treh}
\end{equation}
Then using Eq.~(\ref{xivalue}) and taking $N_e=55$ and $g_*=106.75$ (as a representative value from the Standard Model degrees of freedom), this is
\begin{equation}
    T_{\rm reh}\approx 9.7\cdot 10^{-4}e^{-3N_{\rm reh}/4}M_{\rm Pl}.
    \label{Trehmetric}
\end{equation}
This thermal bath approximately obeys the Bose-Einstein distribution or Fermi-Dirac distribution.  A useful value is the mean energy of the quanta in this thermal distribution, which is 
\begin{equation}
    E_{\rm thermal}\approx 2.7\,T_{\rm reh}
\approx2.6\cdot 10^{-3}e^{-3 N_{\rm reh}/4} M_{\rm Pl}
\end{equation}
(similar for bosons or fermions). As a representative value, if we take $N_{\rm reh}=1$, then this is
\begin{equation}
    E_{\rm thermal}\approx 1.2\cdot 10^{-3}M_{\rm Pl}.
\end{equation}
This thermal energy can be directly compared
to the cut-off energy $E_{\rm max}$ given in Eq.~(\ref{Emaxlambda5}). For moderate $\lambda$, meaning large $\xi$, then $E_{\rm thermal}\gg E_{\rm max}$. This would mean the theory would break down during this early thermal era. On other other hand, when we consider small $\lambda$, the situation improves. These two energies match at
\begin{equation}
    \lambda_m\approx1.3\cdot 10^{-3} \,\,\,\,\,\,(E_{\rm thermal}=E_{\rm max}).
\end{equation}
For $\lambda\ll \lambda_m$ we have $E_{\rm thermal}\ll E_{\rm max}$, so the theory can be directly used during reheating. For the representative value $\lambda=\lambda_5=10^{-5}$, we obtain $E_{\rm thermal}\approx 0.09 \,E_{\rm max}$, an order of magnitude below the cut-off. One could imagine NP entering in this window without radically altering the physics, though the hierarchy between these scales is only moderate.

More details for both during and after inflation are provided in Appendix~\ref{appendixOnShellQuanta}, in which the number of quanta within any given Hubble volume that would violate the unitarity bound is computed and found to be exponentially small for $\lambda\sim 10^{-5}$.

\subsection{Palatini gravity}

In Palatini gravity, the kinetic term in the Einstein frame is slightly simpler
\begin{equation}
\frac{-\mathcal{L}_{\rm kin}}{\sqrt{-g}}=
\frac{1}{2}\frac{1}{1+\xi\phi^2/M_{\rm Pl}^2}  g^{\mu \nu}  \partial_\mu \phi \partial_\nu \phi .
\end{equation}
By transforming to a canonically normalized field $\chi$ and then exploring the inflationary regime, which occurs at large $\chi$, the potential becomes
\begin{equation}\label{VchiPalatini}
    V(\chi)\simeq\frac{\lambda M_{\rm Pl}^4}{4\xi^2}\left(1-8
   e^{-2\sqrt{\xi}\,\chi/M_{\rm Pl}} \right).
\end{equation}
In the slow-roll regime the parameters $\epsilon$ and $\eta$ are approximated by
\begin{align}
    & \epsilon(\chi)=128\xi e^{-4\sqrt{\xi}\,\chi/M_{\rm Pl}}, \\
       & \eta(\chi)=-32\xi e^{-2\sqrt{\xi}\,\chi/M_{\rm Pl}}.
\end{align}
The end of inflation, defined by $\epsilon=1$, is given in this leading approximation by
\begin{equation}
    \chi_{\rm end}={1\over4\sqrt{\xi}}\,M_{\rm Pl}\,\ln(128\xi).
\end{equation}
As in the previous subsection, all that actually matters is that this value is small compared to the field value during inflation, and can hence be ignored.

Carrying out the integral for the number of e-foldings and then inverting to leading order with $N_e\gg1$ gives
\begin{equation}\label{chiPalatini}
  \chi(N_e)={1\over2\sqrt{\xi}}\,M_{\rm Pl}\,\ln\left(32N_e\xi\right).
\end{equation}
The two slow-roll parameters can then be expressed in terms of the number of e-foldings as
\begin{align}
    &\epsilon(N_e)={1\over 8N_e^2\xi},\\
    &\eta(N_e)=-{1\over N_e}.
\end{align}
The prediction for the spectral index and tensor-to-scalar ratio, to leading order, is then
\begin{align}
&    n_s=1-{2\over N_e}\approx0.964,\\
&r_s={2\over N_e^2\xi}\approx {0.0007\over\xi},
\end{align}
where in the last step we evaluated these for $N_e=55$.

Evaluating the variance of density perturbations in the slow-roll inflationary regime gives
\begin{equation}
    \Delta_s={N_e^2\lambda\over12\pi^2\xi}.
\end{equation}
In order for this to match the observed value, the nonminimal coupling $\xi$ must take on a particular value, namely
\begin{equation}
    \xi\approx 4\cdot10^6 N_e^2\lambda.
\label{xiP}\end{equation}
Again using the reference value $\lambda_5=10^{-5}$ and $N_e=55$, this yields
\begin{equation}
\xi\approx 122,000\,\lambda/\lambda_5,
    \end{equation}
with corresponding field value during inflation, from Eq.~(\ref{chiPalatini}), of
\begin{equation}
    \chi\approx 0.028 M_{\rm Pl}\sqrt{\lambda_5/\lambda}\,(1+0.05\,\log(\lambda/\lambda_5)).
\end{equation}
We may compare this value of $\chi$ with the unitary cut-off scale $E_{\rm max} \simeq 4.94 \, M_{\rm Pl}/\sqrt{\xi}$ that we found in Eq.~(\ref{CutPalatini}), 
\begin{equation}
    E_{\rm max}\simeq 0.014\,M_{\rm Pl}\sqrt{\lambda_5/\lambda},
\end{equation}
which is comparable to the result above in metric gravity, Eq.~(\ref{Emaxlambda5}). In this case, the field displacement $\chi$ from the vacuum to the inflationary regime is a factor of $\sim2$ larger than the cut-off scale (for $\lambda\sim 10^{-5}$), challenging the reliability of this EFT.

On the other hand, we can perform a more direct comparison between the thermal energy just after inflation to the cut-off. Here, the energy density at the end of inflation is still close to its value during inflation, i.e., 
\begin{equation}
\rho(t_{\rm end})\approx{\lambda M_{\rm Pl}^4\over 4\,\xi^2}.
\end{equation}
So the reheating temperature is similar to that in Eq.~(\ref{Treh}), but enhanced by a factor of $\sqrt{2}$. When eliminating $\xi$ for $\lambda$ using the condition of Eq.~(\ref{xiP}), this gives an average thermal energy per quanta of
\begin{equation}
    E_{\rm thermal}\approx 2.7\,T_{\rm reh}
\approx1.3\cdot 10^{-4}e^{-3 N_{\rm reh}/4} (\lambda_5/\lambda)^{1/4}\,M_{\rm Pl}.
\end{equation}
For the representative value $N_{\rm reh}=1$, this is
\begin{equation}
    E_{\rm thermal}
\approx6\cdot 10^{-5}(\lambda_5/\lambda)^{1/4}\,M_{\rm Pl}.
\end{equation}
We see that for any reasonable $\lambda$, we have $E_{\rm thermal}\ll E_{\rm max}$. In particular, for $\lambda=\lambda_5=10^{-5}$, we have $E_{\rm thermal}\approx0.004\,E_{\rm max}$, so the thermal energy is comfortably smaller than the cut-off scale.

\subsection{Canonical kinetic model}

In the final simple model the kinetic term in the Einstein frame is taken to be canonical,
\begin{equation}
\frac{-\mathcal{L}_{\rm kin}}{\sqrt{-g}}=
\frac{1}{2} g^{\mu \nu}  \partial_\mu \phi \partial_\nu \phi .
\end{equation}
Hence we do not need any transformation to a new canonically normalized field, but we shall still refer to the field as $\chi=\phi$ for continuity of notation. 
In the slow-roll inflationary regime, which occurs at large $\chi$, the potential becomes
\begin{equation}
    V(\chi)\simeq\frac{\lambda M_{\rm Pl}^4}{4\xi^2}\left(1-
   {2M_{\rm Pl}^2\over\xi\,\chi^2} \right).
\end{equation}
In the slow-roll regime the parameters $\epsilon$ and $\eta$ are approximated by
\begin{align}
    & \epsilon(\chi)={8M_{\rm Pl}^6\over\xi^2\,\chi^6} , \\
       & \eta(\chi)=-{12M_{\rm Pl}^4\over\xi\,\chi^4}.
\end{align}
The end of inflation, defined by $\epsilon=1$, is given in this leading-order approximation by
\begin{equation}
    \chi_{\rm end}={\sqrt{2}\,M_{\rm Pl}\over\xi^{1/3}}.
\end{equation}
As above, what matters is that this is small compared to the field value during inflation, and hence can be ignored.

Carrying out the integral for the number of e-foldings and then inverting to leading order for $N_e\gg1$ yields
\begin{equation}
  \chi(N_e)=
  {2\over\xi^{1/4}}M_{\rm Pl}\,N_e^{1/4}.
\end{equation}
The two slow-roll parameters can then be expressed in terms of the number of e-foldings as 
\begin{align}
    &\epsilon(N_e)={1\over 8N_e^{3/2}\sqrt{\xi}}, \\
    &\eta(N_e)=-{3\over 4N_e}.
\end{align}
The prediction for the spectral index and tensor-to-scalar ratio, to leading order, is then
\begin{align}
&    n_s=1-{3\over 2N_e}\approx0.973 , \\
&r_s={2\over N_e^{3/2}\sqrt{\xi}}\approx {0.005\over\sqrt{\xi}},
\end{align}
where in the last step we evaluated these for $N_e=55$.

Evaluating the variance of density perturbations in the slow-roll inflationary regime gives
\begin{equation}
    \Delta_s={N_e^{3/2}\lambda\over12\pi^2\xi^{3/2}}.
\end{equation}
In order for this to match the observed value, the parameter $\xi$ must take on a particular value, namely
\begin{equation}
    \xi\approx 25,300 N_e\lambda^{2/3}.
\label{xiKin}\end{equation}
As before, using the reference value $\lambda_5=10^{-5}$ and $N_e=55$, gives
\begin{equation}
\xi\approx 645\,(\lambda/\lambda_5)^{2/3},
\end{equation}
with corresponding field value during inflation of
\begin{equation}
        \chi\approx 1.08 M_{\rm Pl} \,(\lambda_5/\lambda)^{1/6}.
\end{equation}
Using the equation for the unitary cut-off scale for this model, $E_{\rm max} \simeq 38 \, M_{\rm Pl} / \sqrt{\xi}$ that we found in Eq.~(\ref{CutCanonical}), this becomes
\begin{equation}
    E_{\rm max}\approx1.5\,M_{\rm Pl}(\lambda_5/\lambda)^{1/3}.
\end{equation}
In this case, for $\lambda\sim 10^{-5}$, the field $\chi$ during inflation is a little smaller than the cut-off scale, suggesting that the use of this effective field theory is at least somewhat reliable.

For the more direct comparison between the thermal energy just after inflation to the cut-off, we note that  the energy density at end of inflation is $\rho(t_{\rm end})\approx \lambda M_{\rm Pl}^4/(4\xi^2)$. Thus, the reheating temperature is approximated by Eq.~(\ref{Treh}) multiplied by a factor of $\sqrt{2}$. Again eliminating $\xi$ for $\lambda$ using the condition of Eq.~(\ref{xiKin}), gives an average thermal energy per quanta of
\begin{equation}
    E_{\rm thermal}\approx 2.7\,T_{\rm reh}
\approx1.7\cdot 10^{-3}e^{-3 N_{\rm reh}/4} (\lambda_5/\lambda)^{1/12}\,M_{\rm Pl}.
\end{equation}
For the representative value $N_{\rm reh}=1$, this is
\begin{equation}
    E_{\rm thermal}
\approx8\cdot 10^{-4}(\lambda_5/\lambda)^{1/12}\,M_{\rm Pl}.
\end{equation}
We see that here for any reasonable $\lambda$, we have $E_{\rm thermal}\ll E_{\rm max}$. In particular, for $\lambda=\lambda_5=10^{-5}$, we have $E_{\rm thermal}\approx0.0005\,E_{\rm max}$, so the thermal energy is significantly smaller than the cut-off.

\section{Unitarity bounds during inflation}\label{sec:UnitarityInflation}

In Sec.~\ref{sec:UnitarityBounds} we have deduced an upper bound $E_{\rm max}$ on the scale $\Lambda_{\rm NP}$ above which NP is necessary to restore unitarity, by considering the scattering of particles around the global minimum of the system, $\langle \phi \rangle \simeq 0$. Together with our estimates in Sec.~\ref{sec:InflationaryObservables}, this implies the existence of NP with a mass \textit{smaller} than the field values during inflation, and hence a dependence of any predictions from the model on this new (though unknown) physics. 

This conclusion, however, is based on the demand for the model to completely describe the dynamics in a regime of small field values $\langle \phi \rangle\simeq 0$. While this condition is crucial for models like Higgs inflation, many inflationary models are only concerned with a relatively small range of field values $\phi\sim \phi_c$. As discussed in Sec.~\ref{sec:unitarity},  it is well-known that the presence of such large background field values for the scalar affects the relevant scattering amplitudes in a way that raises $E_{\rm max}$ parametrically to $\mathcal{O}(M_{\rm Pl}/\sqrt{\xi}$). This suggests that any of the models previously deemed unreliable could, in principle, still provide a good description of the inflationary dynamics as long $E_{\rm max}>\phi_c$ during inflation and the assumptions underlying the analysis relying on $\langle \phi \rangle \simeq 0$ are loosened. While the question whether or not this can be achieved while upholding the general form of the Lagrangian depends entirely on the details of the model beyond the regime of interest,\footnote{At the very least, it appears inevitable that the mapping of the parameters between the UV and IR EFTs would be sensitive to threshold corrections.} unitarity considerations can nevertheless be used to investigate whether or not such a model would be consistent with unitarity, i.e., $E_{\rm max}(\phi\sim \phi_c)> \phi_c$.

In the following subsection, we provide the EFT expansion of the Lagrangian around the inflationary configuration $\langle \phi \rangle \sim \phi_c$ for the multifield model with standard metric gravity, since this was the case with the largest hierarchy $E_{\rm max} < \chi( N_e)$ when scattering amplitudes were calculated around $\langle \phi \rangle  \simeq 0$. We find that the presence of the inflationary background field of Eq.~\eqref{eq:chiofNMetric} indeed raises the unitarity scale to be comparable to, but still smaller than the magnitude of the relevant scalar fields during inflation. We furthermore show in Appendix~\ref{appendixOnShellQuanta} that the scalar condensate representing this classical state can be realized relying entirely on scalar particles with momenta significantly smaller than even the more restrictive upper bound $E_{\rm max}(\phi\simeq 0)$.

Before undertaking the full calculation of scattering amplitudes around the inflationary configuration $\langle \phi \rangle =  \phi_c$, we note that the shift in the unitarity scale $E_{\rm max}$ in this limit is not surprising. During inflation, the inflaton condensate evolves as \cite{Kaiser:2013sna,Schutz:2013fua}
\begin{equation}
\vert \Phi_c \vert^2 \simeq \frac{ 4 N_e}{3} \frac{ M_{\rm Pl}^2}{\xi} .
\label{Phicondensate}
\end{equation}
Given $N_e \lesssim 55$, we therefore have $\xi \vert \Phi_c \vert^2 \gg M_{\rm Pl}^2$ at early times during inflation. In that limit, the effective potential of Eq.~(\ref{eq:Vrescale}) becomes $V (\Phi) \longrightarrow \lambda M_{\rm Pl}^4 / \xi^2$ to leading order, consistent with a free field theory. Similarly, the momentum-dependent operators arising from the noncanonical kinetic sector ${\cal G}_{IJ} (\Phi) \partial_\mu \Phi^I \partial_\nu \Phi^J$ are likewise suppressed. The curvature of the field-space manifold falls rapidly near $\vert \Phi_c \vert$ compared to near $\vert \Phi \vert \sim 0$. In particular, the Ricci curvature scalar of the induced field-space manifold is given by \cite{Kaiser:2012ak}
\begin{equation}
{\cal R} (\Phi) = \frac{2}{ 3 M_{\rm Pl}^2 C^2 (\Phi)} \left[ (1 + 6 \xi)^2 (2 f (\Phi) )^2 - C^2 (\Phi) \right] ,
\label{Riccifs}
\end{equation}
where $2f (\Phi) = M_{\rm Pl}^2 + \xi \vert \Phi \vert^2$ and $C (\Phi) = 2 f (\Phi) + 6 \xi^2 \vert \Phi \vert^2$. In the vicinity of $\vert \Phi_c \vert$ as given in Eq.~(\ref{Phicondensate}), this becomes ${\cal R} (\Phi_c) \longrightarrow 4/ (3 \xi \vert \Phi_c \vert^2)$. In the opposite limit, when we expand the field-space curvature around $\vert \Phi \vert \sim 0$, we find instead ${\cal R} \longrightarrow 8 \xi / M_{\rm Pl}^2$, so the ratio of these (dimensionful) quantities scales as ${\cal R} (\Phi_c) / {\cal R} (\Phi \sim 0) = 1/ ( 8 \xi N_e ) \ll 1$. We therefore expect operators associated with the curved field-space manifold to remain subdominant when expanding the effective action around the inflationary configuration $\langle \phi \rangle \simeq \phi_c$.

\subsection{Parametric effects}

To probe unitarity during inflation, as the condensate evolves as in Eq.~(\ref{Phicondensate}), 
we can perform a field space rotation and expand our fields as $\Phi= \left( \phi^1 , \phi^2, \phi^3 , \phi_c + \phi^4 \right)$, where $\phi_c =  \sqrt{ 4 N_e / 3 \xi} \> M_{\rm Pl}$. Neglecting the potential (assuming $\lambda \ll 1$), the Lagrangian takes the form
\begin{align}
    \frac{-\mathcal{L}_{\rm kin}}{\sqrt{-g}}= & \frac{1}{2}\left( \frac{M_{\rm Pl}^2 \delta_{IJ} }{\bar{M}_{\rm Pl}^2 + \xi |\phi|^2 + 2 \xi \phi_c \phi^4} + \frac{6 M_{\rm Pl}^2 \xi^2  \Phi^M \Phi^N \delta_{IM} \delta_{JN} }{\left( \bar{M}_{\rm Pl}^2 + \xi |\phi|^2+2 \sqrt{\xi} \bar{M}_{\rm Pl} \phi^4 \right)^2 } \right)\partial_\mu \phi^I \partial^\mu \phi^J ,
\end{align}
where we have introduced the auxiliary scale $\bar{M}_{\rm Pl}^2 \equiv M_{\rm Pl}^2 + \xi \phi_c^2$. As before, we can expand these expressions as a series in $\phi$ to obtain the usual EFT form:
\begin{align}
    \frac{-\mathcal{L}_{\rm kin}}{\sqrt{-g}} =& \frac{1}{2} \frac{3}{4 N_e} \delta_{IJ} \partial_\mu \phi^I \partial^\mu \phi^J +  \frac{9 \xi}{4 N_e}  \partial_\mu \phi^4  \partial^\mu \phi^4  \nonumber \\ 
    &+ \frac{9 \xi^2 }{4 N_e \bar{M}_{\rm Pl}^2}  \delta_{IM} \phi^M \partial_\mu \phi^I  \delta_{JN} \phi^N \partial^\mu \phi^J +  \frac{9\xi^{3/2} }{2 N_e \bar{M}_{\rm Pl}} \delta_{IM} \phi^M \partial_\mu \phi^I \partial^\mu \phi^4  \nonumber \\
    &+ \frac{3}{8 N_e} \delta_{IJ} \partial_\mu \phi^I \partial^\mu \phi^J  \sum_{n=1}^\infty \left(- \frac{\xi |\phi|^2 }{\bar{M}_{\rm Pl}^2} - 2 \sqrt{\xi}\frac{\phi^4}{\bar{M}_{\rm Pl}}\right)^n \nonumber \\
    &+  \frac{9 \xi^2 }{4 N_e \bar{M}_{\rm Pl}^2} \left[ \sum_{n=1}^\infty (1+n) \left(- \frac{\xi |\phi|^2 }{\bar{M}_{\rm Pl}^2} - 2 \sqrt{\xi}\frac{\phi^4}{\bar{M}_{\rm Pl}}\right)^n \right] \delta_{IM} \phi^M \partial_\mu \phi^I  \delta_{JN} \phi^N \partial^\mu \phi^J \nonumber \\ 
    &+  \frac{9 \xi^{3/2} }{2 N_e \bar{M}_{\rm Pl}} \left[\sum_{n=1}^\infty (1+n) \left(- \frac{\xi |\phi|^2 }{\bar{M}_{\rm Pl}^2} - 2 \sqrt{\xi}\frac{\phi^4}{\bar{M}_{\rm Pl}}\right)^n \right] \delta_{IM} \phi^M \partial_\mu \phi^I \partial^\mu \phi^4  \nonumber \\ 
    &+ \frac{9 \xi}{4 N_e} \left[ \sum_{n=1}^\infty (1+n) \left(- \frac{\xi |\phi|^2 }{\bar{M}_{\rm Pl}^2} - 2 \sqrt{\xi}\frac{\phi^4}{\bar{M}_{\rm Pl}}\right)^n \right] \partial_\mu \phi^4  \partial^\mu \phi^4 . \label{eq:LInfFull}
\end{align}
The first line suggests introducing new, canonically normalized degrees of freedom $\varphi^I$ as 
\begin{align}
    \varphi^{1,2,3} \equiv \sqrt{\frac{3}{4 N_e}} \phi^{1,2,3} \quad \text{and} \quad \varphi^4 \equiv \sqrt{\frac{3(1+6 \xi)}{4 N_e}} \phi^4.
\end{align}
We can now analyze the interactions in terms of these fields. First, we can identify two separate scales associated with the higher-dimensional operators. The terms in the second line are suppressed by $ \bar{M}_{\rm Pl}/(2 \sqrt{N_e} \xi) \simeq M_{\rm Pl}/(\sqrt{3} \xi)$. The remaining interactions meanwhile are controlled by the factors
\begin{align}
    \frac{\xi |\phi|^2}{\bar{M}_{\rm Pl}^2}\sim \frac{\xi |\varphi^2|}{M_{\rm Pl}^2} \quad \text{and} \quad \frac{\sqrt{\xi} \phi^4}{\bar{M}_{\rm Pl}} \sim \frac{\varphi^4}{M_{\rm Pl}}.
\end{align}
In the regime $\xi \gg 1$ of interest to us these terms are subdominant, suggesting to neglect them for now. The remaining terms in the second line can, however, be absorbed entirely into the kinetic term through another field redefinition
\begin{align}\label{tildephi4}
    \tilde{\varphi}^4 \equiv \varphi^4 + \frac{ \sqrt{3} \xi}{\sqrt{2} M_{\rm Pl}}  \left[(\varphi^1)^2 +(\varphi^2)^2 + (\varphi^3)^2 \ \right] .
\end{align}
In other words, the new three-point interaction precisely cancels out the term that responsible for lowering $E_{\rm max}$ down to $M_{\rm Pl}/\xi$ in the case $\langle \phi\rangle \simeq 0$. The unitarity of the theory is thus controlled by the scale $M_{\rm Pl}/\sqrt{\xi}$, in agreement with previous investigations \cite{Bezrukov:2010jz,Bezrukov:2012hx,Bezrukov:2013fka,Calmet:2013hia}.

\subsection{Precise bound in multifield models with metric gravity}

In the single-field case, it is again useful to move to a canonically normalized scalar $\chi$. When expanding around the inflationary background
\begin{equation}
    \chi=\chi_c+\delta\chi,
\end{equation}
with $\chi_c=\sqrt{3\over2} \,M_{\rm Pl}\, \ln(4N_e/3)$,
we obtain
\begin{align}
    V\approx & V_0 + {\lambda M_{\rm Pl}^4\over N_e\,\xi^2}\sum_{n=1}c_n\left(\delta\chi\over M_{\rm Pl}\right)^n \nonumber \\ 
    =&V_0+\frac{ \sqrt{6} \lambda}{8 \xi^2 N_e} \delta \chi M_{\rm Pl}^3 - \frac{ \lambda}{8 \xi^2 N_e} \delta \chi^2 M_{\rm Pl}^2 + \frac{ \lambda}{12 \sqrt{6} \xi^2 N_e} \delta \chi^3 M_{\rm Pl} - \frac{ \lambda}{144 \xi^2 N_e} \delta \chi^4 + ...
\end{align}
Since the expansion is controlled by the Planck mass, then a simple estimate for the cut-off is $E_{\rm max}\sim M_{\rm Pl}$. While from this pure scalar sector, the actual cut-off can be somewhat higher for small $\lambda$ and large $\xi$, we do not push this further as we know the gravity sector would not allow for a super-Planckian cut-off.

Now we turn to the multifield case. A useful way to organize the degrees of freedom around a large vacuum expectation value (VEV) $\phi_c$ is to organize the field into an SU(2) doublet as
\begin{equation}
\Phi=U\,\left(\begin{array}{c}0\\\phi_c+h\end{array}\right),
\label{Phidoublet}
\end{equation}
with 
\begin{equation}
    U=e^{i\vec{\sigma}\cdot{\vec\pi}/\phi_c},
\end{equation}
where $\sigma^i$ are the Pauli matrices, $\pi^i$ are the Goldstones, and $h$ is the Higgs mode. By series expanding the exponent and using standard properties of the Pauli matrices, the kinetic term becomes
\begin{equation}
    {1\over2}|\partial_\mu\Phi|^2=
    {1\over2}(\partial_\mu h)^2+
    {1\over2}(\partial_\mu \vec{\pi})^2
    +{h\over \phi_c}(\partial_\mu \vec{\pi})^2
    +{h^2\over 2\phi_c^2}(\partial_\mu \vec{\pi})^2+{1\over 6\phi_c^2}\left(
    (\vec{\pi}\cdot\partial_\mu\vec{\pi})^2
    -\vec{\pi}^2(\partial_\mu\vec{\pi})^2\right)
\end{equation}
where we have truncated the expansion at quartic interactions. Note this representation makes it manifest that there are interactions among the Goldstones only if there is more than one Goldstone.

The full Lagrangian is 
\begin{equation}
    {-\mathcal{L}\over\sqrt{-g}}
    ={1\over2\,\Omega^2(h)}|\partial_\mu\Phi|^2
    +{3\xi^2 (\phi_c+h)^2(\partial_\mu h)^2\over
    M_{\rm Pl}^2\,\Omega^4(h)} + V(h),
\end{equation}
where
\begin{equation}
    \Omega^2(h)=1+\xi (\phi_c+h)^2/M_{\rm Pl}^2.
\end{equation}
We can then define canonically normalized Higgs and Goldstone modes as
\begin{equation}
    h_\ca=\alpha\,h,\,\,\,\,\,\vec{\pi}_\ca={\vec{\pi}\over\Omega_0} ,
    \label{hpicanonical}
\end{equation}
where
\begin{equation}
    \Omega_0^2=\Omega^2(0)=1+\xi \phi_c^2/M_{\rm Pl}^2,\,\,\,\,\,\,\alpha=\sqrt{{1\over\Omega_0^2}+{6\xi^2 \phi_c^2\over M_{\rm Pl}^2\Omega_0^4}} .
\end{equation}
To leading order we then have
\begin{align}
    {- \mathcal{L}\over\sqrt{-g}}=&{1\over2}(\partial_\mu h_\ca)^2+{1\over2}(\partial_\mu\vec{\pi}_\ca)^2
    +{h_\ca\over \phi_c\alpha}(\partial_\mu\vec{\pi}_\ca)^2
    +{h_\ca^2\over 2 \phi_c^2 \alpha^2}(\partial_\mu\vec{\pi}_\ca)^2\nonumber\\
   & +{\Omega_0^2\over 6\phi_c^2}\left(
    (\vec{\pi}_\ca\cdot\partial_\mu\vec{\pi}_\ca)^2
    -\vec{\pi}_\ca^2(\partial_\mu\vec{\pi}_\ca)^2\right)
    +{6\xi^2 \phi_c \, h_\ca(\partial_\mu h_\ca)^2\over M_{\rm Pl}^2\Omega_0^4\alpha^3}
    +{3\xi^2  h_\ca^2(\partial_\mu h_\ca)^2\over M_{\rm Pl}^2\Omega_0^4\alpha^4}
    +V(h).
    \label{LGoldstones}
\end{align}
Here we have taken $\Omega=\Omega_0$. 
We note that this expression is valid around any VEV, including the electroweak VEV (and can be used to derive the $E_{\rm max}\simeq 2 M_{\rm Pl}/\xi$ from earlier), but it is of particular use around inflation.
Recall that during inflation the VEV $\phi_c \simeq \sqrt{ 4 N_e / (3 \xi)} \> M_{\rm Pl}$, as in Eq.~(\ref{Phicondensate}). During inflation, with $N_e\gg 1$ and $\xi\gg 1$, one obtains 
\begin{equation}
\Omega^2\approx\Omega_0^2\left(1+\sqrt{2\over3}{h_\ca\over M_{\rm Pl}}+{h_\ca^2\over6\,M_{\rm Pl}^2}\right).
\end{equation}
So this leading-order Lagrangian is only missing Planck-suppressed operators.

We now consider scattering of two Goldstone modes into two Goldstone modes. This can happen via ($\mathcal{A}_{\rm con}$) contact interaction by the first term on the second line of Eq.~(\ref{LGoldstones}), or via ($\mathcal{A}_h$) Higgs exchange by the third term on the first line. The ratio of these two types of amplitudes is parametrically of the order
\begin{equation}
    {\mathcal{A}_{\rm con}\over \mathcal{A}_h}\sim \alpha^2\Omega_0^2.
\end{equation}
Note that around the usual electroweak vacuum $\Omega_0\approx1$ and $\alpha\approx1$, so both types of processes matter. But at large field values relevant for inflation, we have $\alpha^2\Omega_0^2\approx 6\xi\gg1$. This means $\mathcal{A}_{\rm con}\gg \mathcal{A}_h$, so we can ignore the corrections from Higgs exchange when computing Goldstone scattering. 

We note that if we were to gauge the theory and move to unitary gauge, this means that while there are still significant W-boson self-interactions, the coupling to the Higgs is suppressed in the inflationary regime. Famously, it is the coupling to the Higgs that provides unitarity when scattering the longitudinal modes of W-bosons. (See, e.g., Ref.~\cite{Logan:2022uus}.) So we can anticipate this unitarization is now gone, as we now analyze.

The resulting theory for just the Goldstone modes (or equivalently, for the longitudinal modes of the W-bosons) around the inflationary background is then
\begin{equation}
 {-\mathcal{L}_\pi\over\sqrt{-g}}=
{1\over2}(\partial_\mu\vec{\pi}_\ca)^2
+{\Omega_0^2\over 6\phi_c^2}\left(
    (\vec{\pi}_\ca\cdot\partial_\mu\vec{\pi}_\ca)^2
    -\vec{\pi}_\ca^2(\partial_\mu\vec{\pi}_\ca)^2\right).
    \label{GoldstonesEFT}
\end{equation}
From this we can study a process $\pi_1+\pi_2\to\pi_1+\pi_2$. The tree-level scattering amplitude is
\begin{equation}
|\mathcal{A}|={\Omega_0^2\over 2\phi_c^2}E^2(1-\cos\,\theta),
\end{equation}
where $\theta$ is the scattering angle and $E$ is the center of mass energy. From the partial wave expansion $\mathcal{A}=16\pi (a_0+3 a_1\cos\theta)+\ldots$, we identify
$|a_0|=\Omega_0^2E^2/(32\pi \phi_c^2)$. Then the unitarity bound $|a_0|\leq 1/2$ \cite{Logan:2022uus} allows us to deduce the cut-off arising from the $\pi_1+\pi_2\to\pi_1+\pi_2$ scattering alone,
\begin{equation}
E_{\rm max} \approx {4\sqrt{\pi}\,\phi_c\over\Omega_0}.
\end{equation}
At large field values for inflation $\Omega_0\approx \sqrt{\xi}\, \phi_c/M_{\rm Pl}$, implying
\begin{equation}
    E_{\rm max} \approx {4\sqrt{\pi}\,M_{\rm Pl}\over\sqrt{\xi}},
    \label{EmaxMetricInflation}
\end{equation}
where the $\approx$ represents the observation that taking into account further channels can moderately alter (lower) this bound. In particular, one can consider $\pi_1+\pi_1\to\pi_j+\pi_j$ with $j=2,\,3$. Also, in the Standard Model, one can consider the case of $W^+$/$W^-$ scattering into top/anti-top quarks. Parametrically, we find that this amplitude is given by
\begin{equation}
\mathcal{A}(W^+W^-\to t\,\bar{t})\sim
{\Omega_0\,m_t\,E\over \phi_c},
\label{Afermion}
\end{equation}
where $m_t= y_t\,\phi_c/\sqrt{2}$ is the top mass, with $y_t$ the Yukawa coupling. At the value of $E_{\rm max}$ identified in Eq.~(\ref{EmaxMetricInflation}), this amplitude is parametrically $\mathcal{A}(W^+W-\to t\bar{t})\sim y_t$. Since the Yukawa coupling flows to values of order $y_t\sim 0.4$, this appears as a sub-leading contribution, albeit not entirely negligible (one should check the $\mathcal{O}(1)$ factors to be certain).
So the alteration to the cut-off should be relatively modest.

For Higgs inflation with metric gravity, we found in Section~\ref{MetricObs} that we require $\xi = 142$ for $\lambda = 10^{-5}$ to match the amplitude of primordial perturbations consistent with measurements of anisotropies in the cosmic microwave background radiation. For those parameters, the unitarity cut-off scale in Eq.~(\ref{EmaxMetricInflation}) corresponds to $E_{\rm max} \approx 0.6 \, M_{\rm Pl}$. This implies that the backreaction of the scalar field on the NP responsible for restoring unitarity can at best increase the scale where its effects become relevant up to $E_{\rm max}$, independent of the details of this completion.

In Eq.~\eqref{chivaluemetric}, we had found that in a model involving only the scalar degrees of freedom assumed in this analysis, the CMB observables are generated at $\chi_c\approx 5.3 M_{\rm Pl}\Leftrightarrow \phi_c \approx 0.72 M_{\rm Pl}$. Crucially, this implies that for this particular model, even assuming the strongest possible backreaction, such effects would \textit{not} be sufficient for inflation to unfold without the influence of NP to be relevant.

For completeness, we can also compare the cut-off energy to the characteristic scale that is being probed during inflation, which can be taken to the Hubble scale itself $H$. During the inflationary regime, we have $\rho\approx V\approx\lambda M_{\rm Pl}^4/(4\xi^2)$, which implies
\begin{equation}
H_{\rm inf}\approx{\sqrt{\lambda}\,M_{\rm Pl}
\over \sqrt{12}\,\xi}.
\end{equation}
Using Eq.~(\ref{xivalue}) to relate $\xi$ to $\lambda$ and taking $N_e=55$ gives
\begin{equation}
    H_{\rm inf}\approx6\cdot 10^{-6}\,M_{\rm Pl},
\end{equation}
and the cut-off scale is
\begin{equation}
E_{\rm max}\approx
0.6\,M_{\rm Pl}\,(\lambda_5/\lambda)^{1/4}.
\end{equation}
So we see $H_{\rm inf}\ll E_{\rm max}$. In particular, for $\lambda=\lambda_5=10^{-5}$, we have
$H_{\rm inf}\approx10^{-5}E_{\rm max}$: the scale that is being directly probed during inflation is well below the corresponding cut-off.

\section{Conclusions}
\label{sec:Conclusions}

Scalar fields with nonminimal couplings to gravity such as Higgs inflation have been extensively studied in the context of inflationary cosmology. In this article, we have revisited the matter of unitarity violation in such models, for the first time performing a full $S$-matrix analysis. Doing so has enabled us to study the impact of phase space volume factors on the scale of unitarity violation, which were inaccessible in previous investigations relying on scattering amplitudes alone. The corresponding increase in precision enables reliable statements about unitarity in two previously inaccessible regimes: i) single-field models with small self-couplings, and ii) multifield models with small nonminimal couplings. For the former, our method reveals a roughly logarithmic dependence of the scale of unitarity violation on the numerical value of the coupling constant. For small couplings, this implies a restoration of unitarity up to energies $E_{\rm max} \sim {\cal O} (20) \, M_{\rm Pl} / \xi$ for $\lambda \ll 1$. For $\xi \sim {\cal O} (150)$, this scale is significantly above the previous estimate $E_{\rm max}\sim M_{\rm Pl}/\sqrt{\xi}$, which had come from considering scattering around the inflationary condensate in the single-field limit~\cite{Bezrukov:2010jz,Bezrukov:2012hx,Bezrukov:2013fka,Calmet:2013hia,MagninDissertation}. However, this effect is absent in multifield models, in which the unitarity violation is dominated by the interactions arising from a nontrivial field-space metric, for which we find $E_{\rm max} \simeq 2 M_{\rm Pl}/ \xi$, independent of $\lambda$. To illustrate and quantify the impact of these interactions, we have considered a toy-model combining the potential of Higgs inflation with a canonically normalized kinetic sector. 

The $S$-matrix method used in this work allows for the precision necessary to address scenarios with relatively small nonminimal couplings such as critical Higgs inflation \cite{Bezrukov:2014bra}. While the numerical analysis necessary to incorporate phases of ultra-slow-roll in such scenarios lies beyond the scope of this paper, we have provided simple analytical estimates for several models in the slow-roll approximation. As expected, conventional Higgs inflation in metric gravity is only able to reproduce observations for field values 
larger than the scale of unitarity violation. Any predictions arising from this scenario can therefore be expected to depend on the details of the UV completion necessary to restore unitarity. This tension is partially (but not entirely) alleviated in Palatini gravity. Using our toy model, we demonstrate that the unitarity violations can indeed be traced back to the non-diagonal field-space metric. 

We further confirm that the scale at which perturbative unitarity is violated depends strongly on the regime within which one computes the relevant scattering amplitudes. For models like Higgs inflation, which are assumed to describe vastly different energy regimes (during inflation and also near electroweak symmetry-breaking), scattering of particles around the low-energy global minimum of the potential yields $\Lambda_{\rm NP} <E_{\rm max} \ll \phi_c$. On the other hand, for EFTs of inflation that resemble Higgs inflation, but which are designed to describe the inflationary regime alone, a comparable analysis of scattering amplitudes  around the inflaton condensate yields $E_{\rm max} \sim \phi_c$. This result indicates that the tension between the requirement of unitarity and a self-consistent description of the inflationary regime is much reduced.

Our results also show that the requirement of field values significantly larger than the scale of unitarity violation is directly linked to the requirement of achieving a large enough number of e-folds of inflation. This indeed seems to favor models such as critical Higgs inflation, which are capable of producing a relevant number of e-foldings during phases of ultra-slow-roll. 

\section*{Acknowledgments}
We gratefully acknowledge helpful discussions with Elba Alonso-Monsalve, Mikhail Ivanov, Evan McDonough, and Vincent Vennin as well as detailed and extensive discussions with Jos\'{e} R.~Espinosa, Javier Rubio and Tinto~D.~Verano.
M.~P.~H. is supported in part by U.S.~National Science
Foundation grant PHY-2310572. Portions of this work were conducted in MIT's Center for Theoretical Physics and partially supported by the U.S.~Department of Energy under Contract No.~DE-SC0012567. This project was also supported in part by the Black Hole Initiative at Harvard University, with support from the Gordon and Betty Moore Foundation and the John Templeton Foundation. The opinions expressed in this publication are those of the authors and do not necessarily reflect the views of these Foundations.

\appendix

\section{Matrix elements for Scalar Scatterings in Palatini Gravity}\label{sec:MatrixPalatini}

Here, we provide a detailed derivation of the S-matrix element~\eqref{eq:MPal}. We first observe that the form of this equation allows us to factor out several simple-to-deal-with phase space volume factors, reducing the task at hand to the calculation of the amplitude $\mathcal{A}$. For our discussion in Sec.~\ref{sec:UnitarityPalatini}, we are interested in scatterings $\phi^1+\phi^2\to ...$. The corresponding vertices follow from the interaction terms in Eq.~\eqref{eq:DeltaLPalatini2} and~\eqref{eq:DeltaLPalatini},
\begin{align}\label{eq:DeltaLApp}
    \frac{\Delta \mathcal{L}}{\sqrt{-g}} \supset \frac{1}{2} \frac{n!}{k_1! \cdot ... \cdot k_4!}(\phi^1)^{2k_1}\cdot ... \cdot (\phi^4)^{2k_4} g^{\mu \nu} \partial_\mu \phi^I \partial_\nu \phi^I.
\end{align}
These imply in particular that the momenta arising from the kinetic coupling need to be assigned to two particles of the same species, giving rise to two distinct scenarios: i) The momenta are assigned to either $\phi_1$ or $\phi_2$, $I={1,2}$. ii) The momenta are assigned to either $\phi_3$ and $\phi_4$, $I={3,4}$.

We first consider the case $I=1$. The number of outgoing particles $n_1, ... ,n_4$ are related to the exponents $k_1, ... , k_4$ in Eq.~\eqref{eq:DeltaLApp} through
\begin{align}
    n_1=2 k_1+1 , \quad n_2=2 k_2 -1, \quad n_3= 2 k_3, \quad n_4 = 2k_4 .
\end{align}
We can further observe that the total number of outgoing particles $N$ is related to the parameter $n$ in Eq.~\eqref{eq:DeltaLApp} through $N \equiv \sum_i n_i $. This implies that the vertex is given by
\begin{align}
    \mathcal{A}_{I=1}=\frac{\xi^{N/2}}{2 M_{\rm Pl}^{N}} \frac{\left(\frac{N}{2}\right)!}{\left(\frac{n_1-1}{2}\right)!\left(\frac{n_2+1}{2}\right)!\left(\frac{n_3}{2}\right)!\left(\frac{n_4}{2}\right)!} \cdot ...
\end{align}
To obtain the full amplitude we need to factor in permutations of the in- and outgoing particles. The two legs carrying momentum manifest in a factor of $2$, while the remaining legs contribute a factor of $(n_1-1)!(n_2+1)!n_3!n_4!$. Furthermore taking into account all possible distributions of the momenta over the $\phi^1$ particles, the total amplitude is given by
\begin{align}
    \mathcal{A}_{I=1}=&\frac{\xi^{N/2}}{2 M_{\rm Pl}^{N}} \frac{\left(\frac{N}{2}\right)!}{\left(\frac{n_1-1}{2}\right)!\left(\frac{n_2+1}{2}\right)!\left(\frac{n_3}{2}\right)!\left(\frac{n_4}{2}\right)!} \cdot 2 (n_1-1)!(n_2+1)!n_3!n_4! \nonumber \\ 
    &\cdot \left( {q_1}_\mu (p_1^\mu + ... + p_{n_1}^\mu)+ {p_2}_\mu (p_3^\mu + ...  + p_{n_1}^\mu) + ... + {p_{n_1-1}}_\mu p_{n_1}^\mu \right) \sim \nonumber \\ 
    &\sim \frac{\xi^{N/2}}{M_{\rm Pl}^{N}} \frac{ (n_1-1)!(n_2+1)!n_3!n_4! \left(\frac{N}{2}\right)!}{\left(\frac{n_1-1}{2}\right)!\left(\frac{n_2+1}{2}\right)!\left(\frac{n_3}{2}\right)!\left(\frac{n_4}{2}\right)!}  \left( {q_1}_\mu p_1^\mu n_1 + {p_1}_\mu p_2^\mu \frac{n_1}{2}(n_1 -1) \right) .
\end{align}
In the last line, we have collected terms that will turn out to be identical under the LIPS integral and used the simple identity $\sum_{i=1}^k i = k (k+1)/2$. The remaining integrals can then be evaluated using the integral identities~\eqref{eq:LIPSp} and~\eqref{eq:LIPSpp}:
\begin{align}
    \intd & \rho_2 (q) \intd \rho_N (p)\ \mathcal{A}_{I=1} \nonumber  \\ 
    =& \frac{\xi^{N/2}}{M_{\rm Pl}^{N}} \frac{ (n_1-1)!(n_2+1)!n_3!n_4! \left(\frac{N}{2}\right)!}{\left(\frac{n_1-1}{2}\right)!\left(\frac{n_2+1}{2}\right)!\left(\frac{n_3}{2}\right)!\left(\frac{n_4}{2}\right)!} \intd \rho_2 (q) \intd \rho_N (p)\   \left( {q_1}_\mu p_1^\mu n_1 + {p_1}_\mu p_2^\mu \frac{n_1}{2}(n_1 -1) \right)   \nonumber \\ 
    =&  \frac{\xi^{N/2}}{M_{\rm Pl}^{N}} \frac{ (n_1-1)!(n_2+1)!n_3!n_4! \left(\frac{N}{2}\right)!}{\left(\frac{n_1-1}{2}\right)!\left(\frac{n_2+1}{2}\right)!\left(\frac{n_3}{2}\right)!\left(\frac{n_4}{2}\right)!} \left[ \frac{E^2}{2 N} n_1  {\rm Vol}_2 {\rm Vol}_N  + \frac{(N-2)!}{N!} E^2 \frac{n_1}{2} (n_1-1)  {\rm Vol}_2 {\rm Vol}_N  \right]\nonumber \\ 
    =&  \frac{\xi^{N/2}}{2 M_{\rm Pl}^{N}} \frac{ n_1!n_2!n_3!n_4! \left(\frac{N}{2}\right)!}{\left(\frac{n_1+1}{2}\right)!\left(\frac{n_2+1}{2}\right)!\left(\frac{n_3}{2}\right)!\left(\frac{n_4}{2}\right)!} \left(\frac{n_1+1}{2}\right) (n_2+1) \left[ \frac{1}{N} + \frac{n_1-1}{N-1}   \right] E^2 {\rm Vol}_2 {\rm Vol}_N \nonumber \\ 
    =& \frac{\xi^{N/2}}{4 M_{\rm Pl}^{N}} F  (n_1+1) (n_2+1) \left[ 1 + \frac{n_1-1}{N-1}   \right] E^2 {\rm Vol}_2 {\rm Vol}_N .
\end{align}
The auxiliary function $F$ combines several combinatorial factors,
\begin{align}
    F= \frac{n_1! n_2! n_3! n_4! \left( \frac{N}{2} \right)!}{N \left( \frac{n_1+1}{2} \right)!\left( \frac{n_2+1}{2} \right)!\left( \frac{n_3}{2} \right)!\left( \frac{n_4}{2} \right)!}.
\end{align}
This also implies that for $I=2$,
\begin{align}
    \intd \rho_2 (q) \intd \rho_N (p)\ \mathcal{A}_{I=2}=& \frac{\xi^{N/2}}{4 M_{\rm Pl}^{N}} F  (n_1+1) (n_2+1) \left[ 1 + \frac{n_2-1}{N-1}   \right] E^2 {\rm Vol}_2 {\rm Vol}_N .
\end{align}
Next, we consider $I=3$. This corresponds to the vertex with
\begin{align}
    n_1=2 k_1-1 , \quad n_2=2 k_2 -1, \quad n_3= 2 k_3+2, \quad n_4 = 2k_4 .
\end{align}
Using Eq.~\eqref{eq:DeltaLApp}, we can read off the the vertex factor as
\begin{align}
    \mathcal{A}_{I=3}=\frac{\xi^{N/2}}{2 M_{\rm Pl}^N} \frac{ \left( \frac{N}{2} \right)!}{\left( \frac{n_1+1}{2} \right)!\left( \frac{n_2+1}{2} \right)!\left( \frac{n_3}{2} -1 \right)!\left( \frac{n_4}{2} \right)!}\cdot ...
\end{align}
Exchanging the momenta again gives a factor of $2$, while the momentum-independent legs manifest in a factor $(n_1+1)!(n_2+1)!(n_3-2)!n_4!$. Taking into account different distributions of momenta, we ultimately arrive at
\begin{align}
    \mathcal{A}_{I=3}=&\frac{\xi^{N/2}}{M_{\rm Pl}^N} \frac{(n_1+1)!(n_2+1)!(n_3-2)!n_4! \left( \frac{N}{2} \right)!}{\left( \frac{n_1+1}{2} \right)!\left( \frac{n_2+1}{2} \right)!\left( \frac{n_3}{2} -1 \right)!\left( \frac{n_4}{2} \right)!}\cdot \left( {p_1}_\mu (p_2^\mu + ... + p_{n_3}^\mu)+... + {p_{n_3-1}}_\mu p_{n_3}^\mu \right)\nonumber \\ 
    \sim & \frac{\xi^{N/2}}{M_{\rm Pl}^N} (n_1+1) (n_2+1) \frac{n_3}{2} \frac{n_1!n_2!(n_3-2)!n_4! \left( \frac{N}{2} \right)!}{\left( \frac{n_1+1}{2} \right)!\left( \frac{n_2+1}{2} \right)!\left( \frac{n_3}{2} \right)!\left( \frac{n_4}{2} \right)!} \cdot {p_1}_\mu p_2^\mu \frac{n_3}{2} (n_3-1) \nonumber \\ 
    =&  \frac{\xi^{N/2}}{4 M_{\rm Pl}^N} N \cdot F \cdot (n_1+1)(n_2+1) n_3 \cdot {p_1}_\mu p_2^\mu .
\end{align}
Performing the LIPS integral we then find
\begin{align}
    \intd \rho_2 (q) \intd \rho_N (p)\ & \mathcal{A}_{I=3} =  \frac{\xi^{N/2}}{4 M_{\rm Pl}^N} N \cdot F \cdot (n_1+1)(n_2+1) n_3 \intd \rho_2 (q) \intd \rho_N (p)\  {p_1}_\mu p_2^\mu \nonumber \\ 
    =&  \frac{\xi^{N/2}}{4 M_{\rm Pl}^N} N \cdot F \cdot (n_1+1)(n_2+1) n_3 \cdot \frac{(N-2)!}{N!} E^2  {\rm Vol}_2 {\rm Vol}_N  \nonumber \\ 
    =& \frac{\xi^{N/2}}{4 M_{\rm Pl}^N}  F \cdot (n_1+1)(n_2+1) \frac{n_3}{N-1} E^2  {\rm Vol}_2 {\rm Vol}_N  .
\end{align}
The case $I=4$ is now again trivial,
\begin{align}
    \intd \rho_2 (q) \intd \rho_N (p)\ \mathcal{A}_{I=4} = \frac{\xi^{N/2}}{4 M_{\rm Pl}^N}  F \cdot (n_1+1)(n_2+1) \frac{n_4}{N-1} E^2  {\rm Vol}_2 {\rm Vol}_N .
\end{align}
Altogether, this implies that the LIPS integral over the total amplitude is given by
\begin{align}
    \intd \rho_2 (q) \intd \rho_N (p)\ \sum_{I=1}^4 \mathcal{A}_{I} = \frac{\xi^{N/2}}{4 M_{\rm Pl}^N}  F \cdot (n_1+1)(n_2+1) \left( 2+\frac{N-2}{N-1} \right) E^2  {\rm Vol}_2 {\rm Vol}_N .
\end{align}
It is now straightforward to obtain the $S$-matrix element,
\begin{align}
    \mathcal{M}= &\left[n_1! ... n_4! \cdot{\rm Vol}_2 \cdot {\rm Vol}_N  \right]^{- \frac{1}{2}}  \intd \rho_2 \intd \rho_N \  \mathcal{A}_{I=4} \nonumber \\ 
    =&   \frac{E^2\xi^{N/2}}{4 M_{\rm Pl}^N}  \frac{\left[n_1! ... n_4! \cdot{\rm Vol}_2 \cdot {\rm Vol}_N  \right]^{\frac{1}{2}} \left( \frac{N}{2} \right)!}{N \left( \frac{n_1+1}{2} \right)!\left( \frac{n_2+1}{2} \right)!\left( \frac{n_3}{2} \right)!\left( \frac{n_4}{2} \right)!} (n_1+1)(n_2+1) \frac{3N-4}{N-1}  \nonumber \\
    =&  \frac{E^2\xi^{N/2}}{32 \pi M_{\rm Pl}^N} \left(\frac{n_1! ... n_4!}{(N-2)!(N-1)!}\right)^{\frac{1}{2}} \frac{(3N-4) \left( \frac{N}{2} \right)! (n_1+1)(n_2+1)  }{N (N-1) \left( \frac{n_1+1}{2} \right)!\left( \frac{n_2+1}{2} \right)!\left( \frac{n_3}{2} \right)!\left( \frac{n_4}{2} \right)!}\left( \frac{E}{4 \pi} \right)^{N-2} \nonumber \\ 
    =&  \frac{\pi}{2} \left(\frac{n_1! ... n_4!}{(N-2)!(N-1)!}\right)^{\frac{1}{2}} \frac{(3N-4) \left( \frac{N}{2} \right)! (n_1+1)(n_2+1) }{N (N-1) \left( \frac{n_1+1}{2} \right)!\left( \frac{n_2+1}{2} \right)!\left( \frac{n_3}{2} \right)!\left( \frac{n_4}{2} \right)!}\left( \frac{\sqrt{\xi} E}{4 \pi M_{\rm Pl}} \right)^{N} .
\end{align}

\section{Expected Numbers of On-Shell Quanta with $E \sim E_{\rm max}$}
\label{appendixOnShellQuanta}

In Section \ref{sec:UnitarityBounds}, we derived bounds on the energy scale $E_{\rm max}$ above which tree-level perturbative scattering amplitudes among individual on-shell quanta would violate unitarity, when calculating scattering amplitudes around the global minimum $\langle \phi \rangle \simeq 0$. The most constraining values from those analyses yielded the values in Eqs.~(\ref{Cutmetric}), (\ref{CutPalatini}), and (\ref{CutCanonical}):
\begin{equation}
E_{\rm max} \simeq \left\{ \begin{array}{cl}  2 M_{\rm Pl}/\xi & \text{metric gravity} \\  4.9 M_{\rm Pl} /\sqrt{\xi} & \text{Palatini gravity} \\  38 M_{\rm Pl}/\sqrt{\xi} & \text{canonical kinetic.} \end{array} \right.
\label{EmaxCases}
\end{equation}
The values of of $E_{\rm max}$ in Eq.~(\ref{EmaxCases}) came from considering momentum-dependent operators associated with the noncanonical kinetic sectors in multifield models. The values in Eq.~(\ref{EmaxCases}) are each evaluated for $\lambda = 10^{-5}$. Meanwhile, we saw in Sec.~\ref{sec:InflationaryObservables} that $\lambda$ and $\xi$ are not independent; they are (mutually) constrained by the requirement to produce an appropriate amplitude of primordial perturbations on scales probed by the CMB. Upon setting $\lambda = 10^{-5}$, we found $\xi = 142$ (metric gravity), $\xi = 122,000$ (Palatini gravity), and $\xi = 645$ (canonical kinetic sector). 

A relevant question to consider is how likely it would be to find one or more quanta with such on-shell energies within, say, a Hubble sphere during or immediately after inflation. In this appendix we show that although such few-body scatterings would indeed violate unitarity, the likelihood that even one particle within an entire Hubble sphere would have such energy during or after inflation is negligible.

Inflation in such models is driven by a condensate of low-momentum Higgs quanta with $k \ll H$, where $H$ is the Hubble scale during inflation. We consider an equilibrium distribution function for bosons, $f (x, p) \rightarrow n_{\rm B} (k)$, with the usual boson occupation number per mode,
\begin{equation}
n_{\rm B} (k) = \frac{1}{ \exp [ E (\vert {\bf k} \vert ) / T] - 1} ,
\label{nB}
\end{equation}
where we adopt natural units ($\hbar = k_B = c = 1$) and consider Higgs quanta far above the SM electroweak symmetry-breaking scale. In that regime, the Higgs particles are effectively massless, so $E \rightarrow \vert {\bf k} \vert$. Within a few efolds of inflation, we may take the temperature $T$ to be the Gibbons-Hawking temperature associated with the de Sitter horizon \cite{Gibbons:1977mu}:
\begin{equation}
T = T_{\rm dS} = \frac{ H}{2\pi} .
\label{Tds}
\end{equation}
During slow roll, the Friedmann equation (in the Einstein frame) reduces to
\begin{equation}
H^2 \simeq \frac{ V (\Phi )}{3 M_{\rm pl}^2} .
\label{Friedmann}
\end{equation}
(Because we are neglecting contributions from kinetic terms, this slow-roll expression for $H$ in Eq.~(\ref{Friedmann}) holds for all three models of interest.) For most of the duration of inflation, we may evaluate $H$ along the plateau of the Einstein-frame potential, such that $V (\Phi) \simeq \lambda M_{\rm Pl}^4 / (4 \xi^2)$, and hence
\begin{equation}
T_{\rm dS} \simeq {\cal N} \frac{ M_{\rm Pl}}{\xi},
\label{TdsN}
\end{equation}
where we have defined the dimensionless factor
\begin{equation}
{\cal N} \equiv \frac{1}{ 2 \pi} \sqrt{ \frac{\lambda}{12}} = 0.046 \cdot \lambda^{1/2} .
\label{Ndef}
\end{equation}
For realistic values of $\lambda$ at inflationary energy scales, such as $\lambda = 10^{-5}$, we have ${\cal N} = 1.5 \cdot 10^{-4} \ll 1$. The occupation number for Higgs quanta with $E = E_{\rm max}$ would therefore be
\begin{equation}
n_{\rm B} (k = E_{\rm max}) \simeq \left\{ \begin{array}{cl}  \exp[ - 2 / {\cal N}] & \text{metric gravity}\\
\exp [ -4.9 \sqrt{\xi} / {\cal N}] & \text{Palatini gravity} \\
\exp [- 38 \sqrt{\xi} / {\cal N} ] &\text{ canonical kinetic.}
\end{array} \right.
\label{nBEmax}
\end{equation}
Note that for $\lambda = 10^{-5}$ and hence ${\cal N} = 1.5 \cdot 10^{-4}$, we have $1 / {\cal N} \gg 1$; for the latter two cases, this large number is then multiplied by $\sqrt{\xi} \gg 1$ within the exponential term. For the case of multifield Higgs inflation in metric gravity, Eq.~(\ref{nBEmax}) yields $n_{\rm B} (E_{\rm max}) \sim {\cal O} (10^{-5978})$; the occupation number is exponentially smaller for the other two cases, with $n_{\rm B} (E_{\rm max}) <  10^{-2.6 \cdot 10^6}$ for each.

Next we consider the likelihood to find a single on-shell Higgs quantum with $E \sim E_{\rm max}$ within a Hubble sphere during inflation. The number density of particles with energies $E \geq E_{\rm max}$ is
\begin{align}
n (E \geq E_{\rm max}) &= \int_{k_{\rm max}}^\infty \frac{ d^3 k}{(2 \pi)^3} \, n_{\rm B} (k) \nonumber \\
&= \frac{ T_{\rm dS}^3}{2\pi^2} \left[ \left( \frac{ E_{\rm max}}{T_{\rm dS}} \right)^2 e^{- E_{\rm max} / T_{\rm dS}} \right] \left[ 1 + {\cal O} \left( \frac{T_{\rm dS}}{E_{\rm max} } \right) \right] .\label{numdensity}
\end{align}
The number of on-shell Higgs quanta within a Hubble sphere of radius $R_H = H^{-1}$ with $E \geq E_{\rm max}$ during inflation is then given by
\begin{align}
N (E \geq E_{\rm max}) &= \frac{4\pi}{3} \frac{n (E \geq E_{\rm max})}{ (2 \pi T_{\rm dS} )^3} \nonumber  \\
&= \frac{1}{ 12 \pi^4} \left[ \left( \frac{ E_{\rm max}}{T_{\rm dS}} \right)^2 e^{- E_{\rm max} / T_{\rm dS}} \right] \left[ 1 + {\cal O} \left( \frac{T_{\rm dS} } {E_{\rm max} } \right) \right] . \label{NHiggsHubbleSphere}
\end{align}
We find $N (E \geq E_{\rm max}) < 10^{-5974}$ for Higgs inflation in metric gravity, and $N (E \geq E_{\rm max}) < 10^{-2.6\cdot 10^6}$ for the other two cases. In other words, during inflation there are essentially zero Higgs quanta with on-shell energy $E \sim E_{\rm max}$ within a spatial volume exponentially larger than a Hubble sphere.

We likewise expect a scarcity of on-shell quanta with $E \sim E_{\rm max}$ within a Hubble sphere during and after post-inflation reheating \cite{Amin:2014eta,Allahverdi:2020bys}. During the resonant preheating phase, the inflaton condensate oscillates with frequency $\omega \sim \sqrt{\lambda} M_{\rm pl} / \xi \ll E_{\rm max}$ \cite{DeCross:2015uza}, and quanta are produced within resonance bands that typically extend up to $(k \tau)^2 \leq 150$ \cite{DeCross:2016fdz,DeCross:2016cbs,Sfakianakis:2018lzf}, where $\tau = 2 \pi / \omega$, so again we find $k \ll E_{\rm max}$ for $\lambda \sim 10^{-5}$.

Following the resonant preheating phase, the system enters a fully nonlinear regime marked by cascades of power among modes of different wavenumbers $k$, ultimately yielding a thermal spectrum at some large equilibrium temperature $T_{\rm reh}$. Although reheating after Higgs inflation is efficient, it is not instantaneous \cite{Bezrukov:2008ut,Garcia-Bellido:2008ycs,Ema:2016dny,Sfakianakis:2018lzf}. The emergence of thermalization typically requires $N_{\rm reh} \sim 2 - 3$ efolds following the end of inflation in such models \cite{Nguyen:2019kbm,vandeVis:2020qcp,Bettoni:2021zhq,Laverda:2023uqv,Bettoni:2024ixe,Figueroa:2024asq}. Approximating the averaged equation of state during reheating as $p = w \rho$ with $w \sim 0$, we have $\rho (t_{\rm reh}) \simeq \rho (t_{\rm end}) \exp[ - 3 N_{\rm reh}]$. We may then use the Friedmann equation to relate $\rho (t)$ to $H (t)$ at the end of inflation $(t_{\rm end})$ and at the end of reheating ($t_{\rm reh})$, and approximate $H^2 (t_{\rm end}) \simeq V (\phi (t_{\rm end}))/(3 M_{\rm Pl}^2)$. For metric gravity, we have $\phi (t_{\rm end}) \simeq M_{\rm Pl} / \sqrt{\xi}$ \cite{DeCross:2015uza}, giving
\begin{align}
H^2 (t_{\rm end}) \simeq \frac{ \lambda}{48} \frac{ M_{\rm Pl}^2}{\xi^2} ,
\label{Hend}
\end{align}
while for Palatini gravity and the canonical kinetic model, this is increased by a factor of 4.
Meanwhile, once reheating yields a thermal state of $g_*$ relativistic degrees of freedom at temperature $T_{\rm reh}$, we have
\begin{align}
H^2 (t_{\rm reh}) = \frac{1}{ 3 M_{\rm Pl}^2} \left( \frac{ \pi^2}{30} g_* T_{\rm reh}^4 \right) .
\label{Hreh}
\end{align}
Then we arrive at the estimate for the reheating temperature $T_{\rm reh}$ as given in Eq.~(\ref{Treh}). For $\lambda = 10^{-5}$, we find $T_{\rm reh} \ll E_{\rm max}$ for each model under consideration, even for instantaneous reheating (with $N_{\rm reh} = 0$).

We write ${\cal E} = T_{\rm reh} / E_{\rm max}$, such that ${\cal E} \ll 1$.
Repeating the steps as above, we estimate the number of Higgs quanta with $E \geq E_{\rm max}$ to be
\begin{align}
N (E \geq E_{\rm max}) &\simeq \frac{4 \pi}{3} \frac{1}{ H^3 (t_{\rm reh})} \frac{ T_{\rm reh}^3}{2 \pi^2} \left( \frac{ e^{- 1 / {\cal E}}}{ {\cal E}^2} \right) \left[ 1 + {\cal O} ({\cal E}) \right] \nonumber \\
&=\frac{2}{3 \pi} \left( \frac{ 90}{\pi^2 g_*} \right)^{3/2} \left(\frac{ M_{\rm Pl}}{T_{\rm reh}} \right)^3 \left( \frac{ e^{- 1 / {\cal E}}}{ {\cal E}^2} \right) \left[ 1 + {\cal O} ({\cal E}) \right],
\label{Nreh1}
\end{align}
where the bottom line follows upon using Eq.~(\ref{Hreh}) to relate $H (t_{\rm reh})$ to $T_{\rm reh}$. Given the constraints on $\xi$ from CMB observations that we found in Section \ref{sec:InflationaryObservables} ($\xi = 142$ for the model with metric gravity; $\xi = 122,000$ for the model with Palatini gravity; and $\xi = 645$ for the model with canonical kinetic sector), we find $N (E \geq E_{\rm max}) < 1$ for the model with metric gravity for $N_{\rm reh} > 0.63$ e-folds. For more realistic durations of reheating in this model, we find $N (E \geq E_{\rm max}) = 2.3 \cdot 10^{-3}$ ($N_{\rm reh} = 1$), $1.2 \cdot 10^{-16}$ ($N_{\rm reh} = 2$), and $1.3 \cdot 10^{-46}$ ($N_{\rm reh} = 3$). Meanwhile, for the model with Palatini gravity, we find $N (E \geq E_{\rm max}) = 1.7\cdot 10^{-176}$ for instantaneous reheating ($N_{\rm reh} = 0$) and values that are more sharply exponentially suppressed for $N_{\rm reh} > 0$. For the model with canonical kinetic terms, $N (E \geq E_{\rm max}) \ll 10^{-300}$ for $N_{\rm reh} = 0$ and is likewise exponentially smaller for $N_{\rm reh} > 0$. We therefore find essentially zero particles with $E \sim E_{\rm max}$ within a volume spanning multiple Hubble spheres during and after reheating in each of these models.

\bibliographystyle{JHEP}
\bibliography{ref}

\providecommand{\href}[2]{#2}\begingroup\raggedright\begin{thebibliography}{10}

\bibitem{ATLAS:2012yve}
{\scshape ATLAS} collaboration, \emph{{Observation of a new particle in the search for the Standard Model Higgs boson with the ATLAS detector at the LHC}}, \href{https://doi.org/10.1016/j.physletb.2012.08.020}{\emph{Phys. Lett. B} {\bfseries 716} (2012) 1} [\href{https://arxiv.org/abs/1207.7214}{{\ttfamily 1207.7214}}].

\bibitem{CMS:2012qbp}
{\scshape CMS} collaboration, \emph{{Observation of a New Boson at a Mass of 125 GeV with the CMS Experiment at the LHC}}, \href{https://doi.org/10.1016/j.physletb.2012.08.021}{\emph{Phys. Lett. B} {\bfseries 716} (2012) 30} [\href{https://arxiv.org/abs/1207.7235}{{\ttfamily 1207.7235}}].

\bibitem{Bezrukov:2007ep}
F.L.~Bezrukov and M.~Shaposhnikov, \emph{{The Standard Model Higgs boson as the inflaton}}, \href{https://doi.org/10.1016/j.physletb.2007.11.072}{\emph{Phys. Lett. B} {\bfseries 659} (2008) 703} [\href{https://arxiv.org/abs/0710.3755}{{\ttfamily 0710.3755}}].

\bibitem{Rubio:2018ogq}
J.~Rubio, \emph{{Higgs inflation}}, \href{https://doi.org/10.3389/fspas.2018.00050}{\emph{Front. Astron. Space Sci.} {\bfseries 5} (2019) 50} [\href{https://arxiv.org/abs/1807.02376}{{\ttfamily 1807.02376}}].

\bibitem{Birrell:1982ix}
N.D.~Birrell and P.C.W.~Davies, \emph{{Quantum Fields in Curved Space}}, Cambridge Monographs on Mathematical Physics, Cambridge Univ. Press, New York (2, 1984), \href{https://doi.org/10.1017/CBO9780511622632}{10.1017/CBO9780511622632}.

\bibitem{Buchbinder:1992rb}
I.L.~Buchbinder, S.D.~Odintsov and I.L.~Shapiro, \emph{{Effective Action in Quantum Gravity}}, {Taylor and Francis}, {New York} (1992).

\bibitem{Parker:2009uva}
L.E.~Parker and D.~Toms, \emph{{Quantum Field Theory in Curved Spacetime}: {Quantized Field and Gravity}}, Cambridge Monographs on Mathematical Physics, Cambridge University Press, New York (8, 2009), \href{https://doi.org/10.1017/CBO9780511813924}{10.1017/CBO9780511813924}.

\bibitem{Burgess:2009ea}
C.P.~Burgess, H.M.~Lee and M.~Trott, \emph{{Power-counting and the Validity of the Classical Approximation During Inflation}}, \href{https://doi.org/10.1088/1126-6708/2009/09/103}{\emph{JHEP} {\bfseries 09} (2009) 103} [\href{https://arxiv.org/abs/0902.4465}{{\ttfamily 0902.4465}}].

\bibitem{Burgess:2010zq}
C.P.~Burgess, H.M.~Lee and M.~Trott, \emph{{Comment on Higgs Inflation and Naturalness}}, \href{https://doi.org/10.1007/JHEP07(2010)007}{\emph{JHEP} {\bfseries 07} (2010) 007} [\href{https://arxiv.org/abs/1002.2730}{{\ttfamily 1002.2730}}].

\bibitem{Barbon:2009ya}
J.L.F.~Barbon and J.R.~Espinosa, \emph{{On the Naturalness of Higgs Inflation}}, \href{https://doi.org/10.1103/PhysRevD.79.081302}{\emph{Phys. Rev. D} {\bfseries 79} (2009) 081302} [\href{https://arxiv.org/abs/0903.0355}{{\ttfamily 0903.0355}}].

\bibitem{Hertzberg:2010dc}
M.P.~Hertzberg, \emph{{On Inflation with Non-minimal Coupling}}, \href{https://doi.org/10.1007/JHEP11(2010)023}{\emph{JHEP} {\bfseries 11} (2010) 023} [\href{https://arxiv.org/abs/1002.2995}{{\ttfamily 1002.2995}}].

\bibitem{Giudice:2010ka}
G.F.~Giudice and H.M.~Lee, \emph{{Unitarizing Higgs Inflation}}, \href{https://doi.org/10.1016/j.physletb.2010.10.035}{\emph{Phys. Lett. B} {\bfseries 694} (2011) 294} [\href{https://arxiv.org/abs/1010.1417}{{\ttfamily 1010.1417}}].

\bibitem{Lerner:2011it}
R.N.~Lerner and J.~McDonald, \emph{{Unitarity-Violation in Generalized Higgs Inflation Models}}, \href{https://doi.org/10.1088/1475-7516/2012/11/019}{\emph{JCAP} {\bfseries 11} (2012) 019} [\href{https://arxiv.org/abs/1112.0954}{{\ttfamily 1112.0954}}].

\bibitem{Barbon:2015fla}
J.L.F.~Barbon, J.A.~Casas, J.~Elias-Miro and J.R.~Espinosa, \emph{{Higgs Inflation as a Mirage}}, \href{https://doi.org/10.1007/JHEP09(2015)027}{\emph{JHEP} {\bfseries 09} (2015) 027} [\href{https://arxiv.org/abs/1501.02231}{{\ttfamily 1501.02231}}].

\bibitem{Steingasser:2023ugv}
T.~Steingasser and D.I.~Kaiser, \emph{{Higgs potential criticality beyond the Standard Model}}, \href{https://doi.org/10.1103/PhysRevD.108.095035}{\emph{Phys. Rev. D} {\bfseries 108} (2023) 095035} [\href{https://arxiv.org/abs/2307.10361}{{\ttfamily 2307.10361}}].

\bibitem{Steingasser:2024hqi}
T.~Steingasser, \emph{{Higgs criticality in and beyond the SM}}, \href{https://doi.org/10.22323/1.463.0150}{\emph{PoS} {\bfseries CORFU2023} (2024) 150} [\href{https://arxiv.org/abs/2405.02415}{{\ttfamily 2405.02415}}].

\bibitem{Detering:2024vxs}
M.~Detering and T.~You, \emph{{Vacuum Metastability from Axion-Higgs Criticality}},  \href{https://arxiv.org/abs/2412.03542}{{\ttfamily 2412.03542}}.

\bibitem{Hiller:2024zjp}
G.~Hiller, T.~H\"ohne, D.F.~Litim and T.~Steudtner, \emph{{Vacuum stability in the Standard Model and beyond}}, \href{https://doi.org/10.1103/PhysRevD.110.115017}{\emph{Phys. Rev. D} {\bfseries 110} (2024) 115017} [\href{https://arxiv.org/abs/2401.08811}{{\ttfamily 2401.08811}}].

\bibitem{Bezrukov:2014bra}
F.~Bezrukov and M.~Shaposhnikov, \emph{{Higgs inflation at the critical point}}, \href{https://doi.org/10.1016/j.physletb.2014.05.074}{\emph{Phys. Lett. B} {\bfseries 734} (2014) 249} [\href{https://arxiv.org/abs/1403.6078}{{\ttfamily 1403.6078}}].

\bibitem{Bezrukov:2010jz}
F.~Bezrukov, A.~Magnin, M.~Shaposhnikov and S.~Sibiryakov, \emph{{Higgs inflation: consistency and generalisations}}, \href{https://doi.org/10.1007/JHEP01(2011)016}{\emph{JHEP} {\bfseries 01} (2011) 016} [\href{https://arxiv.org/abs/1008.5157}{{\ttfamily 1008.5157}}].

\bibitem{Bezrukov:2012hx}
F.~Bezrukov, G.K.~Karananas, J.~Rubio and M.~Shaposhnikov, \emph{{Higgs-Dilaton Cosmology: an effective field theory approach}}, \href{https://doi.org/10.1103/PhysRevD.87.096001}{\emph{Phys. Rev. D} {\bfseries 87} (2013) 096001} [\href{https://arxiv.org/abs/1212.4148}{{\ttfamily 1212.4148}}].

\bibitem{Bezrukov:2013fka}
F.~Bezrukov, \emph{{The Higgs field as an inflaton}}, \href{https://doi.org/10.1088/0264-9381/30/21/214001}{\emph{Class. Quant. Grav.} {\bfseries 30} (2013) 214001} [\href{https://arxiv.org/abs/1307.0708}{{\ttfamily 1307.0708}}].

\bibitem{Calmet:2013hia}
X.~Calmet and R.~Casadio, \emph{{Self-healing of unitarity in Higgs inflation}}, \href{https://doi.org/10.1016/j.physletb.2014.05.008}{\emph{Phys. Lett. B} {\bfseries 734} (2014) 17} [\href{https://arxiv.org/abs/1310.7410}{{\ttfamily 1310.7410}}].

\bibitem{MagninDissertation}
A.~Magnin, \emph{Higgs inflation and its self-consistency}, Ph.D. Dissertation, \'{E}cole Polytechnique F\'{e}d\'{e}rale de Lausanne (2013).

\bibitem{Aydemir:2012nz}
U.~Aydemir, M.M.~Anber and J.F.~Donoghue, \emph{{Self-healing of unitarity in effective field theories and the onset of new physics}}, \href{https://doi.org/10.1103/PhysRevD.86.014025}{\emph{Phys. Rev. D} {\bfseries 86} (2012) 014025} [\href{https://arxiv.org/abs/1203.5153}{{\ttfamily 1203.5153}}].

\bibitem{Burgess:2007pt}
C.P.~Burgess, \emph{{Introduction to Effective Field Theory}}, \href{https://doi.org/10.1146/annurev.nucl.56.080805.140508}{\emph{Ann. Rev. Nucl. Part. Sci.} {\bfseries 57} (2007) 329} [\href{https://arxiv.org/abs/hep-th/0701053}{{\ttfamily hep-th/0701053}}].

\bibitem{Burgess:2020tbq}
C.P.~Burgess, \emph{{Introduction to Effective Field Theory}}, Cambridge University Press (New York: 2020), \href{https://doi.org/10.1017/9781139048040}{10.1017/9781139048040}.

\bibitem{Brivio:2017vri}
I.~Brivio and M.~Trott, \emph{{The Standard Model as an Effective Field Theory}}, \href{https://doi.org/10.1016/j.physrep.2018.11.002}{\emph{Phys. Rept.} {\bfseries 793} (2019) 1} [\href{https://arxiv.org/abs/1706.08945}{{\ttfamily 1706.08945}}].

\bibitem{Isidori:2023pyp}
G.~Isidori, F.~Wilsch and D.~Wyler, \emph{{The standard model effective field theory at work}}, \href{https://doi.org/10.1103/RevModPhys.96.015006}{\emph{Rev. Mod. Phys.} {\bfseries 96} (2024) 015006} [\href{https://arxiv.org/abs/2303.16922}{{\ttfamily 2303.16922}}].

\bibitem{Logan:2022uus}
H.E.~Logan, \emph{{Lectures on perturbative unitarity and decoupling in Higgs physics}},  \href{https://arxiv.org/abs/2207.01064}{{\ttfamily 2207.01064}}.

\bibitem{Cheung:2007st}
C.~Cheung, P.~Creminelli, A.L.~Fitzpatrick, J.~Kaplan and L.~Senatore, \emph{{The Effective Field Theory of Inflation}}, \href{https://doi.org/10.1088/1126-6708/2008/03/014}{\emph{JHEP} {\bfseries 03} (2008) 014} [\href{https://arxiv.org/abs/0709.0293}{{\ttfamily 0709.0293}}].

\bibitem{Weinberg:2008hq}
S.~Weinberg, \emph{{Effective Field Theory for Inflation}}, \href{https://doi.org/10.1103/PhysRevD.77.123541}{\emph{Phys. Rev. D} {\bfseries 77} (2008) 123541} [\href{https://arxiv.org/abs/0804.4291}{{\ttfamily 0804.4291}}].

\bibitem{Senatore:2010wk}
L.~Senatore and M.~Zaldarriaga, \emph{{The Effective Field Theory of Multifield Inflation}}, \href{https://doi.org/10.1007/JHEP04(2012)024}{\emph{JHEP} {\bfseries 04} (2012) 024} [\href{https://arxiv.org/abs/1009.2093}{{\ttfamily 1009.2093}}].

\bibitem{Pinol:2024arz}
L.~Pinol, \emph{{Effective field theory of multifield inflationary fluctuations}}, \href{https://doi.org/10.1103/PhysRevD.110.L041302}{\emph{Phys. Rev. D} {\bfseries 110} (2024) L041302} [\href{https://arxiv.org/abs/2405.02190}{{\ttfamily 2405.02190}}].

\bibitem{Azhar:2018nol}
F.~Azhar and D.I.~Kaiser, \emph{{Flows into inflation: An effective field theory approach}}, \href{https://doi.org/10.1103/PhysRevD.98.063515}{\emph{Phys. Rev. D} {\bfseries 98} (2018) 063515} [\href{https://arxiv.org/abs/1807.02088}{{\ttfamily 1807.02088}}].

\bibitem{Azhar:2022yip}
F.~Azhar and D.I.~Kaiser, \emph{{Flows into de Sitter space from anisotropic initial conditions: An effective field theory approach}}, \href{https://doi.org/10.1103/PhysRevD.107.043506}{\emph{Phys. Rev. D} {\bfseries 107} (2023) 043506} [\href{https://arxiv.org/abs/2207.08355}{{\ttfamily 2207.08355}}].

\bibitem{Weymann-Despres:2023wly}
G.~Weymann-Despres, S.~Henrot-Versill\'e, G.~Moultaka, V.~Vennin, L.~Duflot and R.~von Eckardstein, \emph{{MSSM inflation revisited: Toward a coherent description of high-energy physics and cosmology}}, \href{https://doi.org/10.1103/PhysRevD.108.023511}{\emph{Phys. Rev. D} {\bfseries 108} (2023) 023511} [\href{https://arxiv.org/abs/2304.04534}{{\ttfamily 2304.04534}}].

\bibitem{Kaiser:2010ps}
D.I.~Kaiser, \emph{{Conformal Transformations with Multiple Scalar Fields}}, \href{https://doi.org/10.1103/PhysRevD.81.084044}{\emph{Phys. Rev. D} {\bfseries 81} (2010) 084044} [\href{https://arxiv.org/abs/1003.1159}{{\ttfamily 1003.1159}}].

\bibitem{Bauer:2010jg}
F.~Bauer and D.A.~Demir, \emph{{Higgs-Palatini Inflation and Unitarity}}, \href{https://doi.org/10.1016/j.physletb.2011.03.042}{\emph{Phys. Lett. B} {\bfseries 698} (2011) 425} [\href{https://arxiv.org/abs/1012.2900}{{\ttfamily 1012.2900}}].

\bibitem{Abu-Ajamieh:2020yqi}
F.~Abu-Ajamieh, S.~Chang, M.~Chen and M.A.~Luty, \emph{{Higgs coupling measurements and the scale of new physics}}, \href{https://doi.org/10.1007/JHEP07(2021)056}{\emph{JHEP} {\bfseries 07} (2021) 056} [\href{https://arxiv.org/abs/2009.11293}{{\ttfamily 2009.11293}}].

\bibitem{Cohen:2021ucp}
T.~Cohen, N.~Craig, X.~Lu and D.~Sutherland, \emph{{Unitarity violation and the geometry of Higgs EFTs}}, \href{https://doi.org/10.1007/JHEP12(2021)003}{\emph{JHEP} {\bfseries 12} (2021) 003} [\href{https://arxiv.org/abs/2108.03240}{{\ttfamily 2108.03240}}].

\bibitem{Greenwood:2012aj}
R.N.~Greenwood, D.I.~Kaiser and E.I.~Sfakianakis, \emph{{Multifield Dynamics of Higgs Inflation}}, \href{https://doi.org/10.1103/PhysRevD.87.064021}{\emph{Phys. Rev. D} {\bfseries 87} (2013) 064021} [\href{https://arxiv.org/abs/1210.8190}{{\ttfamily 1210.8190}}].

\bibitem{Kaiser:2012ak}
D.I.~Kaiser, E.A.~Mazenc and E.I.~Sfakianakis, \emph{{Primordial Bispectrum from Multifield Inflation with Nonminimal Couplings}}, \href{https://doi.org/10.1103/PhysRevD.87.064004}{\emph{Phys. Rev. D} {\bfseries 87} (2013) 064004} [\href{https://arxiv.org/abs/1210.7487}{{\ttfamily 1210.7487}}].

\bibitem{Schutz:2013fua}
K.~Schutz, E.I.~Sfakianakis and D.I.~Kaiser, \emph{{Multifield Inflation after Planck: Isocurvature Modes from Nonminimal Couplings}}, \href{https://doi.org/10.1103/PhysRevD.89.064044}{\emph{Phys. Rev. D} {\bfseries 89} (2014) 064044} [\href{https://arxiv.org/abs/1310.8285}{{\ttfamily 1310.8285}}].

\bibitem{Kaiser:2015usz}
D.I.~Kaiser, \emph{{Nonminimal Couplings in the Early Universe: Multifield Models of Inflation and the Latest Observations}}, \href{https://doi.org/10.1007/978-3-319-31299-6_2}{\emph{Fundam. Theor. Phys.} {\bfseries 183} (2016) 41} [\href{https://arxiv.org/abs/1511.09148}{{\ttfamily 1511.09148}}].

\bibitem{Liddle:1994dx}
A.R.~Liddle, P.~Parsons and J.D.~Barrow, \emph{{Formalizing the slow roll approximation in inflation}}, \href{https://doi.org/10.1103/PhysRevD.50.7222}{\emph{Phys. Rev. D} {\bfseries 50} (1994) 7222} [\href{https://arxiv.org/abs/astro-ph/9408015}{{\ttfamily astro-ph/9408015}}].

\bibitem{Planck:2013jfk}
{\scshape Planck} collaboration, \emph{{Planck 2013 results. XXII. Constraints on inflation}}, \href{https://doi.org/10.1051/0004-6361/201321569}{\emph{Astron. Astrophys.} {\bfseries 571} (2014) A22} [\href{https://arxiv.org/abs/1303.5082}{{\ttfamily 1303.5082}}].

\bibitem{Planck:2018jri}
{\scshape Planck} collaboration, \emph{{Planck 2018 results. X. Constraints on inflation}}, \href{https://doi.org/10.1051/0004-6361/201833887}{\emph{Astron. Astrophys.} {\bfseries 641} (2020) A10} [\href{https://arxiv.org/abs/1807.06211}{{\ttfamily 1807.06211}}].

\bibitem{BICEP:2021xfz}
{\scshape BICEP, Keck} collaboration, \emph{{Improved Constraints on Primordial Gravitational Waves using Planck, WMAP, and BICEP/Keck Observations through the 2018 Observing Season}}, \href{https://doi.org/10.1103/PhysRevLett.127.151301}{\emph{Phys. Rev. Lett.} {\bfseries 127} (2021) 151301} [\href{https://arxiv.org/abs/2110.00483}{{\ttfamily 2110.00483}}].

\bibitem{Amin:2014eta}
M.A.~Amin, M.P.~Hertzberg, D.I.~Kaiser and J.~Karouby, \emph{{Nonperturbative Dynamics Of Reheating After Inflation: A Review}}, \href{https://doi.org/10.1142/S0218271815300037}{\emph{Int. J. Mod. Phys. D} {\bfseries 24} (2014) 1530003} [\href{https://arxiv.org/abs/1410.3808}{{\ttfamily 1410.3808}}].

\bibitem{Allahverdi:2020bys}
R.~Allahverdi et~al., \emph{{The First Three Seconds: a Review of Possible Expansion Histories of the Early Universe}},  \href{https://arxiv.org/abs/2006.16182}{{\ttfamily 2006.16182}}.

\bibitem{Bezrukov:2008ut}
F.~Bezrukov, D.~Gorbunov and M.~Shaposhnikov, \emph{{On initial conditions for the Hot Big Bang}}, \href{https://doi.org/10.1088/1475-7516/2009/06/029}{\emph{JCAP} {\bfseries 06} (2009) 029} [\href{https://arxiv.org/abs/0812.3622}{{\ttfamily 0812.3622}}].

\bibitem{Garcia-Bellido:2008ycs}
J.~Garcia-Bellido, D.G.~Figueroa and J.~Rubio, \emph{{Preheating in the Standard Model with the Higgs-Inflaton coupled to gravity}}, \href{https://doi.org/10.1103/PhysRevD.79.063531}{\emph{Phys. Rev. D} {\bfseries 79} (2009) 063531} [\href{https://arxiv.org/abs/0812.4624}{{\ttfamily 0812.4624}}].

\bibitem{Ema:2016dny}
Y.~Ema, R.~Jinno, K.~Mukaida and K.~Nakayama, \emph{{Violent Preheating in Inflation with Nonminimal Coupling}}, \href{https://doi.org/10.1088/1475-7516/2017/02/045}{\emph{JCAP} {\bfseries 02} (2017) 045} [\href{https://arxiv.org/abs/1609.05209}{{\ttfamily 1609.05209}}].

\bibitem{Sfakianakis:2018lzf}
E.I.~Sfakianakis and J.~van~de Vis, \emph{{Preheating after Higgs Inflation: Self-Resonance and Gauge boson production}}, \href{https://doi.org/10.1103/PhysRevD.99.083519}{\emph{Phys. Rev. D} {\bfseries 99} (2019) 083519} [\href{https://arxiv.org/abs/1810.01304}{{\ttfamily 1810.01304}}].

\bibitem{Nguyen:2019kbm}
R.~Nguyen, J.~van~de Vis, E.I.~Sfakianakis, J.T.~Giblin and D.I.~Kaiser, \emph{{Nonlinear Dynamics of Preheating after Multifield Inflation with Nonminimal Couplings}}, \href{https://doi.org/10.1103/PhysRevLett.123.171301}{\emph{Phys. Rev. Lett.} {\bfseries 123} (2019) 171301} [\href{https://arxiv.org/abs/1905.12562}{{\ttfamily 1905.12562}}].

\bibitem{vandeVis:2020qcp}
J.~van~de Vis, R.~Nguyen, E.I.~Sfakianakis, J.T.~Giblin and D.I.~Kaiser, \emph{{Time scales for nonlinear processes in preheating after multifield inflation with nonminimal couplings}}, \href{https://doi.org/10.1103/PhysRevD.102.043528}{\emph{Phys. Rev. D} {\bfseries 102} (2020) 043528} [\href{https://arxiv.org/abs/2005.00433}{{\ttfamily 2005.00433}}].

\bibitem{Bettoni:2021zhq}
D.~Bettoni, A.~Lopez-Eiguren and J.~Rubio, \emph{{Hubble-induced phase transitions on the lattice with applications to Ricci reheating}}, \href{https://doi.org/10.1088/1475-7516/2022/01/002}{\emph{JCAP} {\bfseries 01} (2022) 002} [\href{https://arxiv.org/abs/2107.09671}{{\ttfamily 2107.09671}}].

\bibitem{Laverda:2023uqv}
G.~Laverda and J.~Rubio, \emph{{Ricci reheating reloaded}}, \href{https://doi.org/10.1088/1475-7516/2024/03/033}{\emph{JCAP} {\bfseries 03} (2024) 033} [\href{https://arxiv.org/abs/2307.03774}{{\ttfamily 2307.03774}}].

\bibitem{Bettoni:2024ixe}
D.~Bettoni, G.~Laverda, A.L.~Eiguren and J.~Rubio, \emph{{Hubble-induced phase transitions: gravitational-wave imprint of Ricci reheating from lattice simulations}}, \href{https://doi.org/10.1088/1475-7516/2025/03/027}{\emph{JCAP} {\bfseries 03} (2025) 027} [\href{https://arxiv.org/abs/2409.15450}{{\ttfamily 2409.15450}}].

\bibitem{Figueroa:2024asq}
D.G.~Figueroa, T.~Opferkuch and B.A.~Stefanek, \emph{{Ricci Reheating on the Lattice}},  \href{https://arxiv.org/abs/2404.17654}{{\ttfamily 2404.17654}}.

\bibitem{Kaiser:2013sna}
D.I.~Kaiser and E.I.~Sfakianakis, \emph{{Multifield Inflation after Planck: The Case for Nonminimal Couplings}}, \href{https://doi.org/10.1103/PhysRevLett.112.011302}{\emph{Phys. Rev. Lett.} {\bfseries 112} (2014) 011302} [\href{https://arxiv.org/abs/1304.0363}{{\ttfamily 1304.0363}}].

\bibitem{Gibbons:1977mu}
G.W.~Gibbons and S.W.~Hawking, \emph{{Cosmological Event Horizons, Thermodynamics, and Particle Creation}}, \href{https://doi.org/10.1103/PhysRevD.15.2738}{\emph{Phys. Rev. D} {\bfseries 15} (1977) 2738}.

\bibitem{DeCross:2015uza}
M.P.~DeCross, D.I.~Kaiser, A.~Prabhu, C.~Prescod-Weinstein and E.I.~Sfakianakis, \emph{{Preheating after Multifield Inflation with Nonminimal Couplings, I: Covariant Formalism and Attractor Behavior}}, \href{https://doi.org/10.1103/PhysRevD.97.023526}{\emph{Phys. Rev. D} {\bfseries 97} (2018) 023526} [\href{https://arxiv.org/abs/1510.08553}{{\ttfamily 1510.08553}}].

\bibitem{DeCross:2016fdz}
M.P.~DeCross, D.I.~Kaiser, A.~Prabhu, C.~Prescod-Weinstein and E.I.~Sfakianakis, \emph{{Preheating after multifield inflation with nonminimal couplings, II: Resonance Structure}}, \href{https://doi.org/10.1103/PhysRevD.97.023527}{\emph{Phys. Rev. D} {\bfseries 97} (2018) 023527} [\href{https://arxiv.org/abs/1610.08868}{{\ttfamily 1610.08868}}].

\bibitem{DeCross:2016cbs}
M.P.~DeCross, D.I.~Kaiser, A.~Prabhu, C.~Prescod-Weinstein and E.I.~Sfakianakis, \emph{{Preheating after multifield inflation with nonminimal couplings, III: Dynamical spacetime results}}, \href{https://doi.org/10.1103/PhysRevD.97.023528}{\emph{Phys. Rev. D} {\bfseries 97} (2018) 023528} [\href{https://arxiv.org/abs/1610.08916}{{\ttfamily 1610.08916}}].

\end{thebibliography}\endgroup

\end{document}